%% file: LO_HF_16Apr2023_PRC_resubmitted.tex
\documentclass[prc,twocolumn,showpacs]{revtex4-2}

\usepackage{amsmath,amssymb,bm,bbm}
\usepackage{graphicx}
\usepackage{multirow}
\usepackage{color,ulem}

\newcommand{\eq}[1]{\begin{equation} #1 \end{equation}}
\newcommand{\ket}[1]{ | #1 \rangle }
\newcommand{\bra}[1]{ \langle  #1 |}
\newcommand{\overlap}[2]{\langle #1 | #2 \rangle}
\newcommand{\elmx}[3]{\langle #1 | #2 | #3 \rangle}
\newcommand{\vect}[1]{\mathbf{#1}}
\newcommand{\vectop}[1]{\hat{\bm{#1}}}

\newcommand{\mpi}{m_{\pi}}
\newcommand{\gA}{g_{\!A}}
\newcommand{\tensorcoupling}[3]{\left\{#1 \otimes #2\right\}_{#3}}

\makeatletter
\def\p@subsection{}
\makeatother

\begin{document}
	

  \title{Cutoff effects in Hartree--Fock calculations at leading order
    of chiral effective field theory} 

  
\author{M. S\'anchez S\'anchez}
\affiliation{CENBG, UMR 5797, Université de Bordeaux, CNRS, F-33170, Gradignan,
France}

\author{Dao Duy Duc}
\affiliation{IPHC, UMR 7178, Université de Strasbourg, CNRS, F-67000, Strasbourg, France}

\author{L. Bonneau}
\email{bonneau@lp2ib.in2p3.fr}
\affiliation{LP2I Bordeaux, UMR 5797, Université de Bordeaux, CNRS, F-33170, Gradignan,
France}

\date{\today}


\begin{abstract}
  
  We explore the effects on nuclear bulk properties of using
  regularization cutoffs larger than the nucleon mass within the
  chiral effective field theory with a power counting that ensures
  order-by-order renormalization in the two-nucleon system. To do
  so we calculate ground-state properties of the $^{16}$O nucleus in the
  Hartree--Fock approach in a basis made up of plane waves confined
  in a cube. We find that regularization cutoff effects manifest
  themselves in two distinct ways: a strong sensitivity to the
  the counter-terms in attractive singular partial waves (related
  to the sign of the corresponding low-energy constant) and to the
  correction for spurious deeply bound states (for high enough cutoffs).
  In fact the latter happens to deprive the Hartree--Fock
  approximation of yielding bound solutions in nuclei. We conclude that,
  when using a leading-order chiral potential in the
  Nogga--Timmermans--van Kolck's power counting (with a regularization
  cutoff higher than the nucleon mass), one cannot produce a
  selfconsistent mean field free of spurious bound-state effects
  that can serve as a reference state for beyond-mean-field
  methods. For high regularization
  cutoffs which yield an attractive $^3S_1$ contact potential, one can
  at best incorporate in the mean-field solution a partial correction
  for spurious bound states. Then the remaining correction has to be
  added to the residual interaction in a treatment beyond the
  Hartree--Fock approximation. In fact a ``full'' correction in the
  $^3D_2$ channel, with energy shifts of the order of or somewhat
  larger than those recommended in [Phys. Rev. C \textbf{103}, 054304
    (2021)], is possible.
  
\end{abstract}


\maketitle

%
%
%
%

\section{Introduction}

Nowadays ab initio nuclear-structure calculations employing
chiral effective field theory (EFT) potentials in the Weinberg power
counting~\cite{Weinberg90_PLB251,Weinberg91_NPB363,Machleidt11_PhysRep503}
are as numerous as successful (see, e.g., the very recent works of
Refs.~\cite{Gysbers19_Nature15,Soma20_PRC102,Hergert20_FrontPhys8,Maris20_arXiv}). 
In contrast very few studies have been carried out in the 
Nogga--Timmermans--van Kolck (NTvK) power counting~\cite{Nogga05_PRC72},
among which we can quote the works by Machleidt
and collaborators~\cite{Machleidt10_PRC81} in nuclear matter using the
Brueckner--Bethe--Goldstone approach, by Song and
collaborators~\cite{Song17_PRC96,Song19_PRC100} in $A=3$ systems
using the Faddeev approach, and by Yang and
collaborators~\cite{Yang21_PRC103,Yang21_arXiv2109.13303v1} in light nuclei up to
$A=16$ using no-core shell-model and coupled-cluster approaches. Moreover
some difficulties appear in $A>4$ nuclei within the NTvK power
counting according to the latter work.

Based on a study of phaseshift variation with the
regularization cutoff $\Lambda$ in various partial
waves, Nogga and collaborators~\cite{Nogga05_PRC72}
argued that the Weinberg power counting does not yield renormalization-group
invariant potentials and developed an alternate scheme in which the
renormalization of the one-pion-exchange potential at leading order
should be carried out nonperturbatively in the low partial waves where
the tensor part is attractive. In particular large cutoff values are
considered beyond the chiral symmetry-breaking scale of the order of 1~GeV
and spurious deeply bound two-nucleon states appear beyond the range
of applicability of chiral EFT. The renormalized potential thus needs
to be corrected before renormalization-group invariance beyond $A=2$ systems
can be studied, which has been done with success in
Refs.~\cite{Song17_PRC96,Song19_PRC100}. However the consideration
of such large values of the momentum cutoff
$\Lambda$ and its interpretation in
effective field theory is debated~\cite{Epelbaum06_arXiv0609037v2,%
  Long16_EPJA25,Epelbaum18_EPJA54,Valderrama19_EPJA55,Epelbaum19_EPJA55,%
  vanKolck20_FrontPhys8}.

In this context we do not mean to dispute either point of view or line of
approach to the problem of renormalization of the two-nucleon potential
but rather to analyze how large regularization cutoffs can affect the
outcome of nuclear-structure calculations. Because in several ab initio
many-body methods the Hartree--Fock approximation serves as a reference
solution or at least as an intermediate step, we explore the behavior
of the Hartree--Fock solution with potentials at leading order of the NTvK
power counting using a rather large regularization cutoff. In particular
we try to understand the difficulties and deficiencies that can emerge
from using a two-nucleon potential obeying the NTvK power counting at
leading order, some of them met in the coupled-cluster calculations
of Yang and collaborators~\cite{Yang21_PRC103} who showed
that the Hartree--Fock ground-state solution at leading order 
in $^{16}$O has to be deformed in order to obtain 
a (correlated) ground-state energy below the $4\,\alpha$ threshold.

This paper is organized as follows. In Sec.~II we present the two-nucleon
potential at leading order of chiral EFT in the Nogga--Timmermans--van Kolck
power counting, including the determination of low-energy constants and
the treatment of the spurious deeply bound states. Then we present in 
Sec.~III our framework to implement this potential in the Hartree--Fock
approximation through a confined plane-wave representation. The
obtained results are shown in Sec.~IV, followed by a discussion in
Sec.~V. Finally conclusions are drawn in Sec.~VI. 

%
%
%
%

\section{Leading-order chiral potential}

At least in the Weinberg's and NTvK power
countings, the leading order (LO) chiral potential is a
two-nucleon operator of class I (according to Henley--Miller isospin
classification). It is accounted for by the sum of the one-pion
exchange potential (long-range part) and a contact potential
(short-range part)
\eq{
\label{eq_VLO}
\hat V_{\rm LO} = \hat V^{(0)}_{1\pi} + \hat V^{(0)}_{\rm ct} \,,
}
where the superscript indicates that leading-order contributions
only should enter the labeled terms. Although their expressions are
well known we recall them in the appendix A for the sake of
defining our notation and normalization conventions.

In this exploratory work we discard the electromagnetic interaction
because its dominant contribution, the Coulomb potential, does not
influence the qualitative conclusions that we reach and would
significantly increase the computing time because of the exchange part 
of the potential when treated exactly (as opposed to the customary
Slater approximation~\cite{Slater51_PR81}). The latter contribution is
attractive and partly counterbalance the repulsive direct
contribution. As was shown in Ref.~\cite{LeBloas11_PRC84}, the overall
Coulomb contribution $E_C$ to the ground-state energy calculated
within the Skyrme--Hartree--Fock--BCS approach is well approximated by
the liquid-drop expression $E_C = 0.73 \, Z^2A^{-1/3} - 1.15\,
Z^2/A$. For example $E_C$ represents of the order of 10\% of
the binding energy in the $^{16}$O nucleus.

The renormalization of the one-pion exchange potential is performed in
relevant partial waves in momentum space with a separabale
regularization function of the form
\eq{
\label{eq_freg}
f_{\rm reg}(k',k) = e^{-(k^{\prime \, 2n} + k^{2n})(\hbar c/\Lambda)^{2n}} \, ,
}
where $\Lambda$ is the regularization cutoff and has the
dimension of an energy.
It cuts off high momenta in the counter term as well as in the
one-pion exchange term of $\hat V_{\rm LO}$. In this work we choose the
frequently used value $n=2$. The low-energy constants (LECs)
appearing in the counter terms of the renormalized partial waves are
fitted to scattering quantities.

To calculate phase shifts in the $NN$ scattering we consider
the $K$-matrix form of the Lippmann--Schwinger equation~\cite{Glockle83}
in partial-wave momentum basis~$\ket{k(LS)JM}$
\begin{align}
& K_{L'L}^{(SJ)}(k',k;q) = V_{L'L}^{(SJ)}(k',k) 
+ \frac{2\mu}{N'_k\hbar^2} \times  \nonumber \\
  & 
\sum_{L''} \mathcal P \int_0^{\infty} dk^{\prime\prime} \,
k^{\prime\prime \,2} \, \frac{V_{L'L''}^{(SJ)}(k',k'') \,
  K_{L''L}^{(SJ)}(k'',k;q)}{q^2 - k^{\prime\prime \,2}} \:.
\label{eq_LS_K}
\end{align}
In this basis $\hbar c \,k$ is the norm of the relative linear-momentum
vector between the two nucleons, $L$ is the quantum number of
the orbital relative motion, $S$ is the total spin and $J$ results
from the coupling of relative orbital and total spin angular
momenta. The isospin quantum number is deduced from the Pauli
principle. Moreover, in Eq.~(\ref{eq_LS_K}), $\mu$ is the reduced mass
of the two-nucleon system, $\displaystyle \mathcal P\int$ denotes a
Cauchy principal-value integral and $N'_k$ is a partial-wave
normalization factor defined by
\begin{align}
  \overlap{k'(L'S')J'M'}{k(LS)JM} = & N'_k \, \frac{\delta(k'-k)}{k'k} \,
  \delta_{L'L}\delta_{S'S} \times \nonumber \\
  & \delta_{J'J}\delta_{M'M} \,.
\end{align}
In Eq.~(\ref{eq_LS_K}) $V_{L'L}^{(SJ)}$ is the renormalized
potential in the $^{2S+1}(L',L)_J$ partial wave.

In practice we solve Eq.~(\ref{eq_LS_K}) by approximating
$\displaystyle \mathcal P\int_0^{\infty}$ with a Gauss--Legendre quadrature
over the interval $[0;k_{\max}]$ with $\hbar c\,k_{\max} = \Lambda + 700$~MeV.
The latter upper bound is found to be a satisfactory compromise between
accuracy and computational speed. Once discretized the
Lippmann--Schwinger equation becomes a linear system.
For $\Lambda$ values up to 1500 MeV, it is sufficient to
use 70 points--the degree of the Legendre polynomial whose roots
yield the momentum mesh by an affine transformation from $[-1;1]$
to $[0;k_{\max}]$.

%
%

\subsection{Power counting}

Let us now recall the essential aspects of Weinberg's 
and NTvK power countings.


The former scheme relies on dimensional analysis. It
predicts a short-range potential at leading order entering
Eq.~(\ref{eq_VLO}) of the form 
\eq{
\elmx{\vect k'}{\hat V_{\rm ct}^{(0)}}{\vect k} = 
C_s + C_t \, \vectop{\sigma}_1 \cdot
\vectop{\sigma}_2 \,,
}
where
$\vect k$, $\vect k'$ are incoming and outgoing relative momenta,
$C_s$ and $C_t$ are constants,
and $\vectop{\sigma}_1$,
$\vectop{\sigma}_2$ are Pauli spin-$\frac12$ operators.
Moreover the one-pion exchange
potential requires, at leading order, renormalization of its
$S$-wave channels only. The above contact terms can thus serve this
purpose and according to the Weinberg prescription, the
renormalization of these partial waves has to be carried out
nonperturbatively.
Generally speaking, depending on the order retained to
  truncate the chiral expansion of the inter-nucleon potential, one
  may have up to $A$-nucleon terms in the potential and the
  nonperturbative renormalization is carried out in few-body
  systems up to $A$ nucleons. Typically at $\rm N^3LO$ two-body and
  three-body terms appear in the chiral
  potential and the LECs are to be fitted on
  some two-nucleon and three-nucleon data.
  Then calculations of observables in a given nucleus proceed by
  using a many-body method 
to solve the eigenvalue
equation for $\hat H = \hat T_{\rm intr} + \hat V_{\rm LO}$,
where $\hat T_{\rm intr}$ is the
kinetic energy operator in the center-of-mass frame of the nucleus.
We consider here the (modest) Hartree--Fock approximation,
which requires to solve a one-body eigenvalue equation
iteratively. This is the first step in so-called beyond-mean-field
calculations such as, for medium-mass or heavier nuclei,
in the coupled-cluster method~\cite{Hagen14_RepProgPhys106,Sun22_PRC106},
the selfconsistent Gorkov--Green's functions
method~\cite{Soma11_PRC84,Soma21_EPJA57}, 
the projected generator coordinate method~\cite{Bally21_PRC103},
the many-body perturbation
theory~\cite{Tichai18_PLB786,Frosini22_EPJA58_62,Frosini22_EPJA58_63,Frosini22_EPJA58_64}
or the in-medium similarity renormalization group
method~\cite{Tsukiyama11_PRL106,Hergert16_PhysRep621,Stroberg17_PRL118}.


In contrast, Nogga, Timmermans and van Kolck~\cite{Nogga05_PRC72}
advocate that, in addition to the two $S$ waves in the Weinberg power
counting, at least low-$L$, spin-triplet partial waves in which the 
tensor part of the one-pion exchange potential is attractive
should be nonperturbatively renormalized. This is prompted by the
singular nature of this potential. As this is the case in an infinite
number of channels one could fear that infinitely many low-energy
constants are necesseary, depriving the theory of any predictive
power. However, beyond some value of $L$, the centrifugal barrier is
expected to provide enough repulsion to counterbalance the attractive
tensor potential and produce spurious bound states beyond EFT momentum
range of applicability. In practice counter-terms in at least
$^3\!P_0$, $^3\!P_2$ and $^3\!D_2$ partial waves should be promoted to
leading order and treated nonperturbatively according to
Ref.~\cite{Nogga05_PRC72}. A detailed analysis of the centrifugal
suppression taming the attractive, non singular one-pion-exchange
potential has been performed in peripheral spin-singlet partial waves
by Pav{\'o}n Valderrama and collaborators in
Ref.~\cite{PavonValderrama17_PRC95}. This was translated into a
power-counting demotion of the one-pion-exchange potential in singlet
channels. However a similar study has not been carried out to date in
spin-triplet partial waves where the tensor part of $\hat V_{1\pi}$ is
attractive. The only related works that we are aware of are those by
Birse~\cite{Birse06_PRC74}, in the chiral limit of vanishing pion
mass, and by Wu and Long~\cite{Wu19_PRC99}.

In the NTvK scheme, the $\hat V_{1\pi}$ potential in partial waves
other than $^1S_0$, $(^3S_1,^3\!D_1)$ (which corresponds to $^3S_1$,
$^3\!D_1$ and $\varepsilon_1 = {^3S_1-^3\!D_1}$ and the hermitean
conjugate), $^3P_0$, $^3\!P_2$ and $^3\!D_2$ are subleading and should 
be treated perturbatively because for large enough cutoff
$\Lambda$, the iteration of the one-pion exchange potential can become 
large and introduce cutoff dependence beyond the error of truncation
at leading order. However for $\Lambda$ below 2000~MeV or so this
should not happen according to Ref.~\cite{Nogga05_PRC72}. This is the
approach followed in the recent no-core shell-model calculations of
Ref.~\cite{Sanchez20_PRC102}. Instead we adopt here a more
``conservative'' strategy by setting to 0 the LO potential in all
partial waves other than those listed above. We call minimal scheme
this approach and denote it by $\rm NTvK_{\min}$.

Moreover to improve
the leading-order description of the $^1\!S_0$ phase shift in $np$
scattering, we follow the work of Ref.~\cite{Sanchez18_PRC97} by
taking into account the low-momentum scale corresponding to the zero
amplitude through a di-baryon auxiliary field in the effective
Lagrangian. However this produces an energy-dependent potential, easy
to manipulate in a two-nucleon system but much less so in a
many-nucleon system. This is why an on-shell equivalent
momentum-dependent potential has been developed in
Ref.~\cite{Sanchez20_PRC102} and abbreviated DBZ in 
Ref.~\cite{Yang21_PRC103}
\eq{
\label{eq_DBZ}
  V_{\rm ct}^{(^1\!S_0)}(k',k) = 4\pi\,
  \frac{(\hbar c)^3}{m_Nc^2} \, 
\bigg[\frac 1 {\Delta_1} \, \mathcal F\Big(\frac{\hbar c\, k'}{\gamma}\Big) 
  \, \mathcal F\Big(\frac{\hbar c \,k}{\gamma}\Big) + \frac 1
  {\Delta_2}\bigg]\,,
}
where the function $\mathcal F$ is defined by
$\mathcal F(x) = (1+x^2)^{-1/2}$ and $m_N$ is the nucleon mass.

In appendix B we give the results of the
fitting of low-energy constants in the $^1S_0$, $^3S_1$, $^3P_0$,
$^3P_2$ and $^3D_2$ channels.

%
%

\subsection{Treatment of spurious deeply bound states}

In this work we consider three values of $\Lambda$,
  namely 500, 1000 and 1500~MeV. Whereas no spurious bound state is
  found for $\Lambda = 500$~MeV, one is supported in the $^3P_0$
  channel for $\Lambda = 1000$~MeV, and two appear for $\Lambda =
  1500$~MeV in the coupled $(^3S_1,^3D_1)$ and the uncoupled $^3P_0$
  channels. Their energies are reported in table~\ref{tab_boud_states}.

\begin{table}[h]
  \caption{Spurious bound-state (negative) energies in
    the coupled $(^3S_1,^3D_1)$ channel and the uncoupled $^3P_0$,
    $^3D_2$ channels as a function of the regularization cutoff
    $\Lambda$. No bound state is obtained in the other channels for
    the considered cutoff values.\label{tab_boud_states}} 
  \centering
  \begin{tabular}{ccccc}
    \hline \hline
    \multirow{2}{*}{$\Lambda$ (MeV)} &&
    \multicolumn{3}{c}{Bound-state energies (MeV)} \\
    \cline{3-5}
    && $(^3S_1,^3D_1)$ & $^3P_0$ & $^3D_2$ \\
    \hline
    500 && $-$ & $-$ & $-$ \\
    \hline
    1000 && $-$ & $-418.15$ & $-$ \\
    \hline
    1500 && $-2066.48$ & $-234.83$ & $-1352.39$ \\
    \hline \hline
  \end{tabular}
\end{table}

\begin{figure}[h]
  \hspace*{-0.25cm}
  \includegraphics[width=0.5\textwidth]{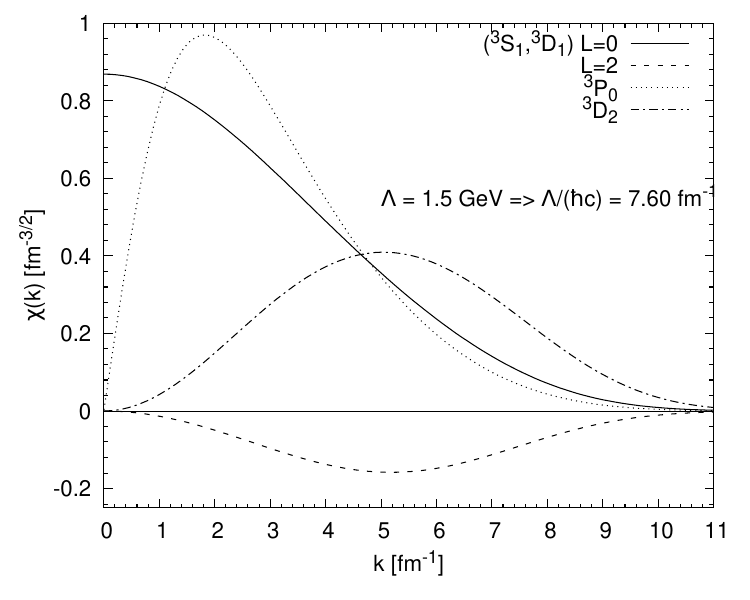}
  \caption{Momentum-space wavefunctions of the spurious deeply
    bound state in the $(^3S_1,^3D_1)$, $^3P_0$ and $^3D_2$ channels for
  $\Lambda = 1500$~MeV as functions of $k$.\label{fig_spurious_wf}}
\end{figure}

We obtain these bound states using the Lagrange-mesh
method~\cite{Baye15_PhysRep565} in momentum space in each partial-wave
channel with the Lagrange--Legendre mesh over the interval $[0;k_{\max}]$
where $\hbar c \,k_{\max} = \Lambda + 700$~MeV as for solving the
Lippmann--Schwinger equation. For $\Lambda$ values up to 1500 MeV,
a 70-point Lagrange--Legendre mesh has been checked to be largely
sufficient. The wave function in momentum space $\chi_i^{(X)}(k)$
of the bound state $\ket{\chi_i}$ in the channel $X$ is real and normalized
to unity as
\eq{
  \frac 1 {N'_k}\sum_{L \in X}\int_0^{\infty}dk \: k^2 \:
  \Big[\chi_{i,L}^{(X)}(k)\Big]^2 = 1 \,,
}
where the sum over relative orbital momentum $L$ is restricted to
the values relevant to the channel $X$. In our implementation
the bound-state wave functions are tabulated at mesh points and
interpolated using the cubic Hermitean spline method 
Ref.~\cite{Huber97_FBS22} frequently used in few-body calculations.
In addition, for low momenta between 0 and the lowest one of the
mesh, we use a parabolic extrapolation from the first three
tabulated values (at the lowest-three mesh points). The resulting
wavefunctions in momentum space are plotted in
figure~\ref{fig_spurious_wf} for the regularization
cutoff $\Lambda=1500$~MeV. 
Only one spurious bound state is supported
in the considered partial waves, so their wave function have no node.
For larger values of $\Lambda$, additional deep bound states
appear~\cite{Song17_PRC96} and one can expect that their wave
functions have an increasing number of nodes.

Once the bound-state wavefunctions are known, the removal of the spurious bound states $\ket{\chi_i}$ from the two-nucleon potential
$\hat V$ can be done by adding a scaled projector on theses bound states as in
Ref.~\cite{Nogga05_PRC72}. The corrected potential $\hat V'$ thus
reads
\eq{
  \hat V' = \hat V + \sum_i E_{\rm shift}^{(i)} \, \ket{\chi_i}\bra{\chi_i}
}
where the energy shifts $E_{\rm shift}^{(i)}$ are large, positive
constants in relevant partial-wave channels generically
numbered~$i$. According to Yang and
collaborators~\cite{Yang21_PRC103},  
sufficiently large values are of the order of 10 to 15~GeV.
We have checked that for any value of $E_{\rm shift}^{(i)}$ the scattering
properties calculated by solving
the Lippmann--Schwinger equation using the above numerical method
with the potential $\hat V'$ (in partial waves supporting spurious
bound states) are unchanged. This is true for any cutoff value $\Lambda$.
Because we work in the momentum partial-wave basis, this treatment
of spurious bound states and the scattering benchmark is straightforward
and efficient. The Lagrange-mesh method in partial-wave representation
is thus a powerful alternative to the harmonic-oscillator basis as
implemented in the NCSM calculations of Ref.~\cite{Yang21_PRC103}.
It is worth mentioning that Wendt and collaborators developed an
alternative method to decouple spurious deeply bound states in the SRG 
framework~\cite{Wendt11_PRC83}, which allows for a softening
of the two-nucleon potential for many-body calculations at the same time
but at the price of inducing three-body forces and higher.

\begin{figure}[h]
  \hspace*{-0.5cm}
  \includegraphics[width=0.525\textwidth]{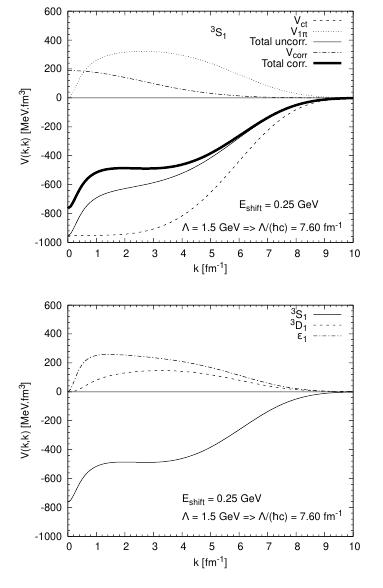}
  \caption{Top panel: Diagonal matrix elements of the corrected potential
    in the $^3\!S_1$ channel as functions of relative momentum $k$
    for $\Lambda=1500$~MeV. The contact $V_{\rm ct}$ and
    one-pion-exchange $V_{1\pi}$ contributions are plotted with
    regularization, as dashed and dotted lines respectively.
    The energy-shift applied to the spurious bound states is 1~GeV as
    an example. Lower panel: 
    Diagonal matrix elements of the potential corrected for
    spurious bound states in the $^3D_1$ and $\varepsilon_1$ channels,
    compared with those in the $^3S_1$ channel, as functions of
    $k$.\label{fig_V3SD1_corrected}} 
\end{figure}

\begin{figure*}[t]
  \includegraphics[width=\textwidth]{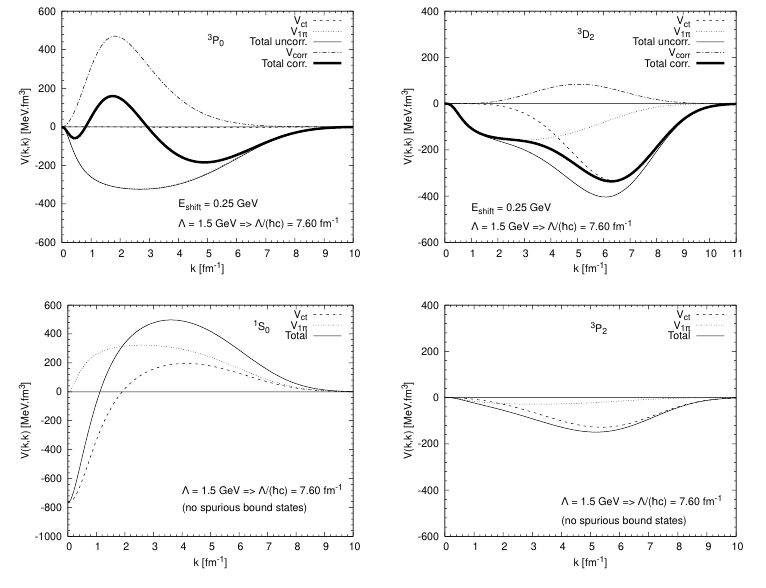}
  \caption{Same as top panel of figure~\ref{fig_V3SD1_corrected}
    in the $^3P_0$ and $^3D_2$ channels (top row) and
    in the $^1S_0$ and $^3P_2$ channels (bottom row).
    No spurious bound-state correction is necessary for the latter channels.
\label{fig_V3P0_V3D2_corrected}}
\end{figure*}

In figures~\ref{fig_V3SD1_corrected} and \ref{fig_V3P0_V3D2_corrected}
we show momentum-space diagonal matrix elements of the corrected potential
$\hat V'$ in the $(^3S_1,^3D_1)$, $^3P_0$ and $^3D_2$ channels
for $\Lambda = 1500$~MeV and an energy shift $E_{\rm shift}^{(i)} = 1$~GeV as an
example. Because the $L=0$ and $L=2$ components of the spurious
bound-state wavefunction in the deuteron channel have constant
and opposite signs as functions of $k$, the projector contributions
to the diagonal channels $^3S_1$ and $^3D_1$ are repulsive whereas
it is attractive in the coupled partial-wave channel $\varepsilon_1$.
Similarly the momentum wavefunction of the $^3P_0$ bound state having
zero node, it is of constant sign and the corresponding contribution
to the corrected potential is repulsive. From these plots it is clear
that the larger is the energy shift, the more repulsive is the corrected
potential in diagonal channels, the relevant ones at leading order for
$\Lambda$ up to 1500~MeV being $^3S_1$ and $^3P_0$. However it is
remarkable that, for $\Lambda = 1500$~MeV, the corrected $^3P_0$-projected
potential is virtually equal to the energy-shifted projector
as the contribution from the counter-term virtually vanishes.
This is a highly non trivial effect of the nonperturbative renormalization
in the $^3P_0$ channel.

For comparison purposes and later discussion, we also display in
figure~\ref{fig_V3P0_V3D2_corrected} the diagonal matrix
elements of $\hat V_{\rm LO}$ in the $^1S_0$ and $^3P_2$
channels as functions of relative momentum. It is worth noting
the repulsive contribution at moderate and high momenta
(for $k$ above 1~$\rm fm^{-1}$)
introduced in the $^1S_0$ channel by the di-baryon formalism and the fit of
its LECs to the amplitude zero.

%
%
%
%

\section{Hartree--Fock approximation in a confined 
plane-wave basis}

To implement the Hartree--Fock approximation to nuclear bound states,
we represent the single-particle states in a basis made of plane
waves confined in a cube, of edge length $L$, centered
at the center-of-mass of the nucleus, to which we add the nucleon spin
projection $\sigma = \pm \frac 1 2$ on a chosen axis (here the $z$
axis) and the isospin projection $\tau = \pm \frac 1 2$.
This representation was first considered by van Dalen and
M\"uther~\cite{vanDalen14_PRC90} (at least in Nuclear Physics) and
more recently studied in details in
Refs.~\cite{Dao20_PhD,Dao20_APhysPolBSupp13}. Here we summarize in
this section its essential aspects. 

The implementation of the confined plane-wave basis to represent the
two-body matrix elements of the most general class-I, II and III
potential and to solve the Hartree--Fock equations resulted in the
so-called \texttt{HFchiral} code. All numerical details and algorithms
can be found in Ref.~\cite{Dao20_PhD}. 
We may put forward three main advantages of this code:
\begin{itemize}
  \item[(i)] it allows to easily describe non spherical shapes,
    especially triaxial ones;
  \item[(ii)] as shown in Appendix~C, when the cubic box is
    large enough, the two-body matrix elements of the potential are
    proportional to those in between relative momenta.
    No recourse to any transformation from the center-of-mass frame to
    the laboratory frame is thus necessary;
  \item[(iii)] there is a direct relation between the cutoffs on
    single-particle and relative momenta. 
    Indeed, following van Dalen and M\"uther~\cite{vanDalen14_PRC90},
    the single-particle basis size, for a fixed edge length $L$, is
    controlled by a truncation on the norm of the momentum
    vector. This truncation scheme is rotationally invariant and thus
    preserves the octahedral symmetry of the basis. The corresponding 
    single-particle momentum cutoff is denoted by $k_{\max}$ and, if
    the potential matrix elements $\elmx{\vect k_{\alpha'}}{\hat V}{\vect
      k_{\alpha}}$ had a sharp cutoff $\Lambda/(\hbar \,c)$ on
    the norm of relative momenta $\vect k_{\alpha}$ and $\vect
    k_{\alpha'}$, we would then also have a sharp cutoff $k_{\max} =
    \Lambda/(\hbar \,c)$ on single-particle momenta.
\end{itemize}
Of course a more efficient implementation of the Hartree--Fock
equations is possible when restricting calculations to spherically
symmetric solutions, as in the ground state of $^{16}$O. However, the
\texttt{HFchiral} code is meant to be used in deformed nuclei.

In the present work we restrict the nuclear shapes to triaxial ones by
building a basis of reducible co-representation of the full octahedral
double group with time-reversal symmetry $\hat{\mathcal T}$,
denoted by $O_{\rm 2h}^{\rm  DT}$. This symmetry-adapted basis allows
to describe time-reversal 
symmetry breaking by the mean field, as in odd-mass nuclei. However,
because the confined plane-wave basis (see Appendix C for its
construction and interpretation) is invariant by
$\hat{\mathcal T}$, it is also
possible to describe time-reversal invariant solutions exactly
($\hat{\mathcal T}$ is then a so-called selfconsistent symmetry).

The full symmetry group of plane waves confined in a cube being a
subgroup of SU(2), our Hartree--Fock calculations break spherical  
symmetry which nevertheless can be approximately recovered as the box
size $L$ increases. Moreover the $j=\frac 1 2$ and $j=\frac 3 2$
irreducible representations of SU(2), of dimensions 2 and 4, 
respectively, are each decomposed in a single irreducible
representation of $O^{\rm DT}_{\rm 2h}$. This means that for nuclei up to $N=8$
and/or $Z=8$ the spherical symmetry breaking of the mean field can be
tamed. Indeed, our Hartree--Fock
ground-state solutions $\ket{\Phi}$ in $^{16}$O are such 
that $\elmx{\Phi}{\hat J_z}{\Phi} = 0$, where $\hat J_z$ is the
component along the $z$ axis of the total angular-momentum operator
$\vectop J$. However we do not expect that $\elmx{\Phi}{\vectop
  J^2}{\Phi} = 0$. To estimate the deviation from exact spherical
symmetry, we calculate the hexadecapole moments
$\elmx{\Phi}{\hat{Q}_{4\mu}}{\Phi}$ and $\elmx{\Phi}{r^4}{\Phi}$. As
was shown in \cite{Dudek07_IJMPE16,Dudek10_JPG37} the relations 
\begin{subequations}
  \begin{align}
    \label{Q2}
    \elmx{\Phi}{\hat{Q}_{\lambda\mu}}{\Phi} & = 0 \quad \mbox{$\forall \, \mu
    \in \{-\lambda,\cdots,\lambda\}$ and $\lambda = 2$, 3} \\ 
    \label{Q40}
  \elmx{\Phi}{\hat{Q}_{44}}{\Phi} & =
  \elmx{\Phi}{\hat{Q}_{4\,-\!4}}{\Phi} =
  -\sqrt{\frac5{14}} \, \elmx{\Phi}{\hat{Q}_{40}}{\Phi} \\
    \label{Q4}
  \elmx{\Phi}{\hat{Q}_{4\mu}}{\Phi} & = 0
  \quad \forall \, \mu \in \{\pm1, \pm2,\pm 3\}
\end{align}
\end{subequations}
are characteristic of a solution $\ket{\Phi}$ with purely octahedral
symmetry, as far as hexadecapole moments are concerned (similar
relations hold for higher multipoles). Therefore the deviation from
the spherical shape in such a solution is essentially encoded in the
Bohr axial-hexadecapole deformation parameter $\beta_4$, which can be
approximated to first order as 
\eq{
  \beta_4 \approx \frac{4\pi}7 \,
  \frac{\elmx{\Phi}{\hat{Q}_{40}}{\Phi}}%
       {\elmx{\Phi}{r^4}{\Phi}} \,.
}
In this context, provided the box size parameter $L$ is large enough,
we expect a weak lowering of the ground-state energy if
angular-momentum projection is performed after the Hartree--Fock
calculation. This corresponds to calculations where a
small deviation $|\beta_4|$ from spherical shapes occurs, typically
less than about 0.01.

In a finite discrete orthonormal basis, the Hartree--Fock equations,
resulting from the variational principle applied to a Slater determinant
trial wave function~\cite{Ring-Schuck80}, takes the usual form of an
eigenvalue equation for the single-particle Hartree--Fock Hamiltonian 
$\hat h_{\rm HF}$. This one-body operator is defined by
its matrix elements between any single-particle states $\ket{a}$ and
$\ket{b}$ (containing all quantum numbers)
\eq{
  \elmx{a}{\hat h_{\rm HF}}{b} = \Big(1-\frac 1 A\Big) \,
  \elmx{a}{\frac{\hat{\vect p}^2}{2m_N}}{b} +
  \sum_{i \in \Phi} \elmx{ai}{\hat V + \hat T_2}{\widetilde{bi}}
}
where $i$ labels all single-particles states occupied in the Slater
determinant $\ket{\Phi}$, $\hat T_2$ is the two-body
correction to the kinetic energy
$$
\hat T_2 = -
\sum\limits_{1\leqslant \mu<\nu \leqslant A}
\frac{\hat{\vect p}_{\mu} \cdot
  \hat{\vect p}_{\nu}}{m_N}
$$
and $\ket{\widetilde{bi}} = \ket{bi} - \ket{ib}$
accounts for direct and exchange contributions, respectively, to the
two-body matrix element. The factor $1/A$ results from the one-body
kinetic-energy correction because at leading order neutrons and protons
have the same mass. The kinetic-energy operator $\hat T_{\rm intr}$ in
the center-of-mass frame (also called intrinsic kinetic energy) thus reads
\eq{
  \hat T_{\rm intr} = \Big(1-\frac 1 A\Big) \sum_{\mu=1}^A
  \frac{\hat{\vect p}_{\mu}^2}{2m_N} - \sum\limits_{1\leqslant \mu<\nu \leqslant A}
\frac{\hat{\vect p}_{\mu}\cdot \hat{\vect p}_{\nu}}{m_N} \,.
}

Since the occupied states $\ket{i}$ are eigenstates
of $\hat h_{\rm HF}$
\eq{
  \label{eq_HF_eq}
\hat h_{\rm HF}\ket i = e_i\,\ket i
}
where $e_i$ denotes the single-particle energy, the eigenvalue equation
Eq.~(\ref{eq_HF_eq}) is solved iteratively, from an initial one-body
potential chosen to be a Woods--Saxon potential with central and spin
orbit contributions. Owing to symmetries imposed to $\hat h_{\rm HF}$,
parity $p$, $z$-signature $r_z$ and isospin projection
$\tau$ are good quantum numbers
in the present work, so the Hartree--Fock equations~(\ref{eq_HF_eq})
are solved independently for each of the four combinations of $(p,r_z)$
for neutrons and protons separately. In nuclei with $N \ne Z$ the
single-particle Hartree--Fock Hamiltonian is different for neutrons
and protons even if the potential $\hat V$ is of class I (which is
the case here at leading order for the strong interaction
and in the absence of the electromagnetic interaction).

%
%
%
%

\section{Results}

We use the \texttt{HFchiral} code developed by two of the authors
(see Refs.~\cite{Dao20_PhD,Dao20_APhysPolBSupp13}) along the lines of
section~III to calculate ground-state bulk properties of the
$^{16}$O nucleus with the leading-order two-nucleon potential
described in section~II.

As explained in the previous section two
parameters characterize our truncated single-particle basis, namely
the edge length $L$ of the confining cubic box and the single-particle
momentum cutoff $k_{\max}$. The former establishes a natural infra-red (IR)
momentum scale $\lambda_{\rm IR} = \frac{\pi}L$, whereas the latter
directly defines an ultra-violet (UV) relative-momentum cutoff
$\lambda_{\rm UV} = k_{\max}$. In order to obtain UV convergence one
should choose $\lambda_{\rm UV}$ somewhat larger than the regularization
cutoff $\Lambda$ of the NN potential, whereas large $L$ values
guarantee IR convergence. Contrary to the spherical
harmonic-oscillator (SHO) basis, the IR and UV momentum scales are
independent of each other in the confined plane-wave basis. 

To get an estimate of a ``reasonable'' value for $L$ and to reach a
compromise between accuracy and computation resources, we
assume the nuclear mean-field potential to have the form of a
Woods--Saxon potential 
\eq{
V_{\rm WS}(r) = - \frac{V_0}{1+ e^{(r-R)/a}}
}
where $R = R_0A^{1/3}$ is the empirical nuclear radius, with $R_0
\approx 1.35$~fm, $V_0$ is the depth of the potential and $a$ the
diffuseness parameter. The box size $L$ should be large enough to
cover the range of an eigenfunction of the corresponding hamiltonian,
namely 
\eq{
  \label{eq_L}
L = 2R + \eta_a \times 2a + \eta_E \times \lambda_E \,.
}
In this expression $\lambda_E = \frac{\hbar c}{\sqrt{-2mc^2\,E}}$ is
the binding wave length (inverse of the binding momentum) and
$E$ is the energy of the last occupied nucleon (Fermi level) in the
Woods--Saxon Hamiltonian neglecting spin-orbit. In the form (\ref{eq_L})
of $L$, the first two contributions give the radius $R_{\rm WS}$ at
which the Woods--Saxon potential can be considered to vanish. Then for
$r> R_{\rm WS}$ the wavefunction of the bound state exponentially
decays according to its eigenenergy $E <0$, with a characteristic
length $\lambda_E$. The third contribution to $L$ in Eq.~(\ref{eq_L})
thus corresponds to the distance beyond $R_{\rm WS}$ at which the
bound-state wave function can be considered to vanish.
The numerical parameters $\eta_a$ and $\eta_E$ are of the order of a few
units so that the single-particle wave function can be considered to
be approximately zero on the edge of the box. Taking $V_0 = 70$~MeV
and $a = 0.7$~fm as typical values, we find a Fermi energy $E =
-32.7$~MeV. We have checked that choosing $\eta_a = \eta_E = 3$ is
enough to converge all bound states in this potential. This results in
$L = 13.4$~fm. As shown by numerical results below, IR
  convergence will be actually reached for slightly smaller values.

%
%

\subsection{Results for a low regularization cutoff}

We begin with the regularization cutoff $\Lambda = 500$~MeV which
corresponds to a somewhat soft NN potential. This enables us to avoid
the spurious-bound state problem and to thoroughly probe the IR and UV
convergences with our available computing resources.
\begin{figure}[h]
  \hspace*{-0.25cm}
  \includegraphics[width=0.525\textwidth]{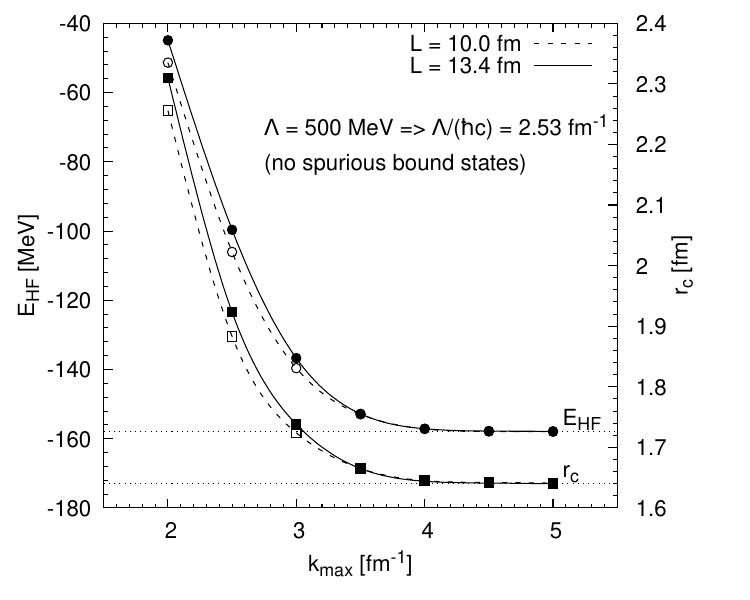}
  \caption{Ground-state energy $E_{\rm HF}$ (circles, left scale) and charge
    radius $r_c$ (squares, right scale) in the Hartree--Fock
    solution of $^{16}$O as functions of the single-particle
    momentum cutoff $k_{\max}$ with $\Lambda = 500$~MeV for two values
    of the cubic box size, $L=10$~fm (dashed line to guide the eye and
    open symbols) and $L=13.4$~fm (solid line and filled symbols).
    \label{fig_HF_O16_Lambda=500}} 
\end{figure}

In Fig.~\ref{fig_HF_O16_Lambda=500} we plot the ground-state
energy and the charge radius as functions of $k_{\max}$ for $L =
10.0$~fm (dashed line and open symbols) and $L =
13.4$~fm (full line and filled symbols).
The dotted lines represent the converged values of $E_{\rm HF}$ and
$r_c$. One clearly observe UV convergence for $k_{\max} \gtrsim 3.5$~$\rm
fm^{-1}$, somewhat higher than the $NN$ potential regularization
cutoff $\Lambda/(\hbar c) \approx 2.53$~$\rm fm^{-1}$. Moreover the
converged value for each observable is the same at the scale of the
figure for both box sizes, which shows IR convergence.
Finally we checked that, for both box sizes $L=10$~fm and
$L=13.4$~fm, the expectation values of the hexadecapole moments in
the Hartree--Fock ground-state solutions $\ket{\Phi}$
obey the relations (\ref{Q2}) to (\ref{Q4}). Moreover we find
$\beta_4$ values of are of the order of a few thousandths.

%
%

\subsection{Results for a high regularization cutoff}

First we use a regularization cutoff $\Lambda = 1000$~MeV.
Disregarding the spurious bound-state problem, we obtain a positive
ground-state energy. Although the total energy is well converged, this
solution is not physical as it corresponds to a discretized unbound
solution. Correlatively the charge radius is very large. In fact 
it is found to oscillate between two distinct values larger than
5~fm. We checked that the same solution is reached when starting from
an extremely deep and wide Woods--Saxon potential. This unbound
solution results from a too strongly repulsive contact potential in
the $^3S_1$ channel. Moreover removing the effect of the $^3P_0$
spurious bound state is expected to yield even more repulsion. 
Therefore in the rest of this subsection we do not further
consider results with $\Lambda = 1000$~MeV and
perform only calculations with $\Lambda = 1500$~MeV.

\begin{figure}[h]
  \includegraphics[width=0.525\textwidth]{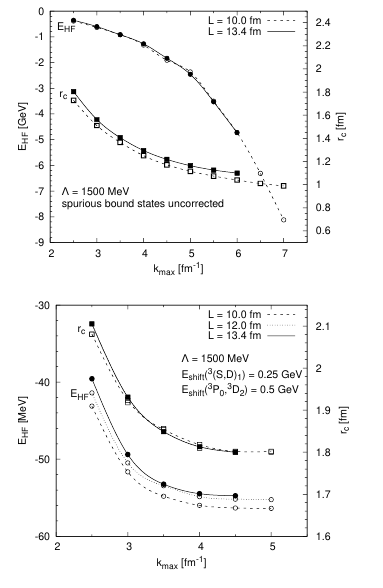}
  \caption{Ground-state energy $E_{\rm HF}$ in the Hartree--Fock
    solution of $^{16}$O as a function of the single-particle
    momentum cutoff $k_{\max}$ with $\Lambda = 1500$~MeV for various
    box sizes $L$ without (top panel)
    and with (bottom panel) correction for spurious bound
    states. Symbols show calculated values, whereas the curves serve
    to guide the eye.
    \label{fig_HF_O16_kmax}}
\end{figure}

Using $\Lambda = 1500$~MeV and disregarding first the spurious bound-state
correction, we obtain the ground-state energy and charge radius 
plotted as functions of the UV cutoff $k_{\max}$ in the top panel of
Fig.~\ref{fig_HF_O16_kmax}. 
With a ``small'' box of edge length $L=10$~fm, the ground-state energy and
the charge radius are found to be very close to the corresponding
values obtained with $L=13.4$~fm, which establishes IR convergence at
least up to $k_{\max} =6$~$\rm fm^{-1}$, the maximum single-particle
cutoff that our computation resources allow us to consider with the
``large'' box. Using the small box allows us to push further
calculations, up to $k_{\max} =7$~$\rm fm^{-1}$. However this value of
$k_{\max}$ is not enough to exhibit UV convergence. Finally we
checked that, as for the low-regularization cutoff $\Lambda =
500$~MeV, our Hartree--Fock ground-state solutions $\ket{\Phi}$
obtained with
both box sizes $L=10$~fm and $L=13.4$~fm, satisfy Eqs.~(\ref{Q2})
to (\ref{Q4}). Moreover, with the smaller box of edge length
$L=10$~fm, the values of $\big|\beta_4\big|$ are sizable, of the
order of 0.1 to 0.2, while they are of the order of a few
hundredths with the larger box of edge length $L=13.4$~fm. We
attribute this difference to a smaller value of
$\elmx{\Phi}{r^4}{\Phi}$ obtained with the smaller box, similarly to
the slightly smaller charge-radius values for $L=10$~fm than with
$L=13.4$~fm as can be seen on the top panel of
Fig.~\ref{fig_HF_O16_kmax}. Therefore we can expect much smaller
values of $\beta_4$ when correcting for spurious bound states, even
partly, because the repulsion introduced by the corrective terms
should increase $\elmx{\Phi}{r^4}{\Phi}$ (as well as the charge
radius).

Let us now address the correction of spurious bound states in the
two-nucleon potential. The Hartree--Fock ground-state solution
should not be bound if the energy shifts are too large, and it
turns out that the recommended values of Ref.~\cite{Yang21_PRC103} are
far above the maximal values yielding bound Hartree--Fock
solutions. Moreover bound solutions obtained with spurious
bound-states correction are expected to be very sensitive to the
energy shifts. In this context our twofold goal is (i) to find a
choice of simultaneous energy shifts for which IR and UV convergence
are reached and (ii) to probe the response of the Hartree--Fock
ground-state energy and charge radius to the energy shift in each
channel supporting a spurious bound state.

Such a possible set of energy shifts is, for example,
$E_{\rm shift}(^3(S,D)_1)=0.25$~GeV and $E_{\rm
  shift}(^3P_0)=E_{\rm shift}(^3D_2)=0.5$~GeV, and yields the results
shown in the bottom panel of Fig.~\ref{fig_HF_O16_kmax} as functions
of $k_{\rm max}$ for three different box sizes. As can be seen, the
repulsion brought by the spurious bound-state correction provides UV
convergence for $k_{\max} \gtrsim 4$~$\rm fm^{-1}$. Moreover IR
convergence is again reached for edge lengths somewhat larger than
$L=10$~fm. It is therefore appropriate to use the single-particle
basis parameters $k_{\max} = 4$~$\rm fm^{-1}$ and $L=10$~fm to study
the variation of the ground-state energy and the charge radius with
the energy shift in each of the $^3(S,D)_1$, $^3P_0$ and $^3D_2$
channels, around the above set of $E_{\rm shift}$ values. The
corresponding results are displayed in the left panels of
Fig.~\ref{fig_HF_O16_Eshift}, whereas the right panels show the
partial-wave contributions to the nuclear interaction energy.
Overall the sensitivity of $E_{\rm HF}$ and $r_c$ to the energy shift
is decreasing as the relative orbital-angular momentum increases. This
is consistent with the fact that, in theses observables, $S$-wave
contributions to the NN potential dominate over the $P$-wave
contributions, which themselves are larger than the $D$-waves. This
hierarchy is observed in phenomenological effective
``interactions'' such as those of the Skyrme type, and seems to
reflect the centrifugal suppression at work in two-nucleon
phaseshifts. It is worth noting that, in the $^3D_2$ partial wave, it
is even possible to apply energy shifts as large as 100~GeV without
getting unbound solutions. Therefore, it is possible to completely
correct for the spurious bound state in the $^3D_2$ channel using
recommended values of Ref.~\cite{Yang21_PRC103}. However the
Hartree--Fock ground-state solution does not converge as a function of
$E_{\rm shift}(^3D_2)$. Indeed if the strength $E_{\rm shift}$ of the
corrected $^3D_2$ potential tends to $+\infty$, then its
one-body reduction (the mean-field potential) becomes increasingly
repulsive for any finite one-body density, so the single-particle
states become unbound and the iterative process tend to diverge.

Another general observation from Fig.~\ref{fig_HF_O16_Eshift} is that,
regardless of the channel in which the energy shift is applied,
all partial-wave contributions to the nuclear interaction energy
$E_{\rm int}$ decrease in absolute value and, with the exception of
$^3S_1$, even seem to converge with $E_{\rm shift}$. Therefore the non
convergence of the Hartree--Fock solution with $E_{\rm shift}$ seems
to be due to the $^3S_1$ channel of the NN potential. It is worth
noting that despite its strong repulsive matrix elements, the $^1S_0$
NN potential gives a sizable negative contribution to $E_{\rm int}$
(hence attraction) as the energy shift increases beyond some value
which depends on the spurious-bound-state channel. This is a highly
nonlinear effect resulting from the Hartree--Fock procedure.

Finally it is worth mentionning that our partially
corrected results in Fig.~\ref{fig_HF_O16_kmax} all correspond to
octahedral-symmetric solutions, obeying Eqs.~(\ref{Q2}) to
(\ref{Q4}), with very small $\beta_4$ values of the order of a few
thousandths for all considered box sizes. Therefore the ground-state
shapes are very close to spherical ones.

%
%
%
%


\section{Discussion}

From the above results we conclude that the energy shift tends to suppress
nuclear binding in the Hartree--Fock solution in addition to
introducing a strong dependence on the energy shift. In fact we can even
expect that no bound solution can be obtained at the Hartree--Fock
level when using recommended values of energy-shift parameters
$E_{\rm shift}^{(i)}$ (10 to 15~GeV according to Ref.~\cite{Yang21_PRC103}).

This situation as at variance with the ab initio many-body methods
based on the diagonalization of the nuclear Hamiltonian $\hat H$, such
as the No-Core Shell Model. Up to truncation effects, solving the
eigenvalue equation of $\hat H' = \hat H + \sum_i E_{\rm shift}^{(i)}
\ket{\chi_i}\bra{\chi_i}$ yields low-energy solutions approximately
independent of $E_{\rm shift}^{(i)}$ (positive) values provided they are large
enough. A residual dependence is expected because of the truncation of
the underlying one-body harmonic-oscillator basis and the many-body
basis. According to the Ritz variational principle, the energy
functional is stationary around the eigenstates of $\hat H$ if one
works in the full Hilbert space. Therefore one expects that, if the
trial wavefunction of the variational principle was rich enough, one
would get a $E_{\rm shift}^{(i)}$-independent result for the ground state if
$E_{\rm shift}^{(i)}$ values are large enough to decouple the spurious bound
states from the physical ones. The extreme sensitivity of the
Hartree--Fock ground-state solution can thus be attributed to its
Slater determinant form, in other words to its breaking of translation
symmetry and the lack of beyond-mean-field correlations (in particular
those associated with the restoration of broken symmetries).
This clearly shows that not only the Hartree--Fock approximation, and
related approaches like the Hartree--Fock--Bogoliubov approach (which
variationally incorporates one-body effects of pairing correlations),
intrinsically break renormalization-group invariance, but they cannot
even provide a reference solution for beyond mean-field calculations
if one tries to implement a full correction for spurious bound-state
effects.
\begin{figure*}[t]
  \includegraphics[width=0.85\textwidth]{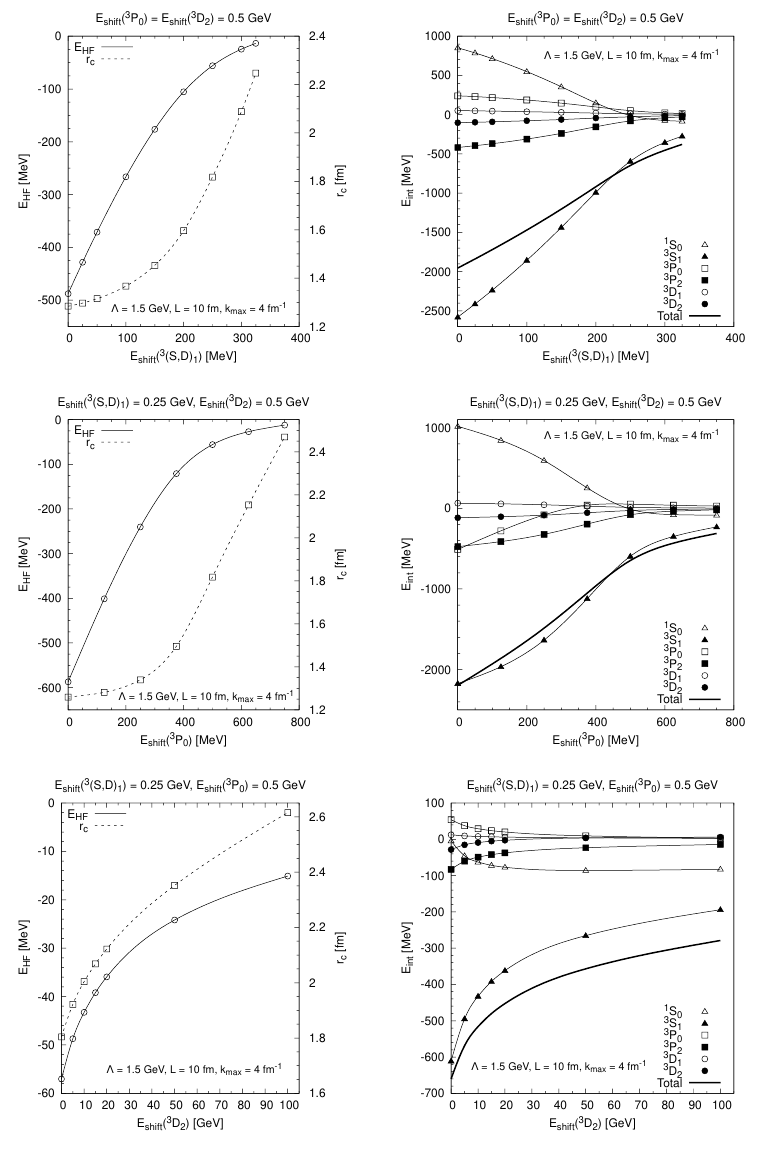}
  \caption{Left panels: ground-state energy $E_{\rm HF}$ (open circles, left scale)
    and charge radius (squares, right scale) in the Hartree--Fock
    solution of $^{16}$O as a function of the energy shift in the
    $^3(S,D)_1$ (top row), $^3P_0$ (middle row) and $^3D_2$ (bottom
    row) channels with a box size $L = 10$~fm and a single-particle
    momentum cutoff $k_{\max} = 4$~$\rm fm^{-1}$. Right panels:
    interaction energy $E_{\rm int} = \elmx{\Phi}{\widehat
      V_{\rm NN}}{\Phi}$ as a function of the energy shift in the
    corresponding channels. \label{fig_HF_O16_Eshift}}
\end{figure*}

A possible way out in mean-field based approaches is
\clearpage
\noindent
to split the spurious-bound-state correction term into a contribution
treated at the mean-field level and the remaining one treated together
with the residual interaction beyond mean field. This allows to bring
in enough repulsion to effectively ``soften'' the inter-nucleon
potential and yield a UV convergence of mean-field results for
tractable single-particle bases, without compromising the
bound character of the ground-state solutions. We can even correct ``fully''
for the $^3D_2$ spurious bound states by employing energy shifts of
the order of 15 to 20~GeV as recommended by Yang and
collaborators~\cite{Yang21_PRC103}. Indeed such large values of
$E_{\rm shift}$, and even somewhat larger ones, do not prevent from
obtaining bound Hartree--Fock solutions as shown in the last row of
Fig.~\ref{fig_HF_O16_Eshift}. However one has to keep in mind
that, as shown by the results with the $\Lambda = 1$~GeV
regularization cutoff, mean-field bound solutions can be obtained only
for $\Lambda$ values for which the contact potential is
attractive. Because the $^3S_1$ channel is dominant, this corresponds
essentially to $C_{^3S_1}(\Lambda) < 0$. 

%
%
%
%

\section{Conclusions}

After renormalizing the one-pion-exchange potential
in the Nogga--Timmermans--van Kolck power counting at leading order,
we attempted Hartree--Fock calculations in $^{16}$O with
a regularization cutoff $\Lambda$ up to 1500~MeV. The single-particle
wave functions and the potential matrix elements are represented in a
basis of plane waves confined in a cubic box of edge length several times
larger than the nuclear radius and truncated according to the norm of
the single-particle momentum. The largest relative momentum retained in
the potential matrix elements thus identifies with this single-particle
momentum cutoff.

In this framework we first highlighted the strong sensitivity of the
Hartree--Fock ground-state solution to the regularization cutoff $\Lambda$
as a direct consequence of the running of the low-energy constants in
partial-wave channels where the one-pion-exchange potential is attractive 
and singular (exhibiting a limit-cycle-like behavior). In particular
we identified the $^3S_1$ counter-term as being dominantly responsible
for either an extremely attractive or a strongly repulsive renormalized
potential. In addition to this behavior, the two-nucleon renormalized
potential can yield spurious deeply bound states in the two-nucleon
system, which requires a correction in order to remove the unphysical
excess of attraction above some value of $\Lambda$. This is done by
adding to the potential the projectors on the spurious bound states
with a weight playing the role of an energy shift. We analyzed the effect
on the Hartree--Fock ground-state solution of the corrected potential
as a function of the energy shift and the partial-wave channel. We
found a weak dependence on the $^3D_2$ energy shift, making
so-corrected Hartree--Fock calculations meaningful. However, in stark
contrast, even a modest energy shift in the $^3S_1$ or $^3P_0$ channels
completely suppresses nuclear binding in the Hartree--Fock solution.

From this study we conclude that one cannot build a Hartree--Fock
solution free of spurious-bound-state effects that can serve as a
reference state for many-body approaches such as coupled cluster
or in-medium SRG. The treatment of the spurious bound states
has thus to be implemented at least partly beyond Hartree--Fock. We
showed that a partial treatment of the corrected two-nucleon potential
at the Hartree--Fock level can be done with a single-particle basis
with a momentum cutoff significantly below the regularization cutoff
$\Lambda$ of the potential. This makes feasible such partly corrected
calculations even for high values of $\Lambda$. However the remaining
part of the corrected two-nucleon potential has to be treated
beyond the Hartree--Fock approximation and this requires high momentum
truncations, which is extremely challenging (see, e.g.,
Ref.~\cite{Yang21_arXiv2109.13303v1}). 

\begin{acknowledgments}
  We thank U. van Kolck, M. Pav\'on Valderrama and C.-J. Yang for
  valuable discussions. Part of computer time for this study was provided
  by the computing facilities MCIA (M\'esocentre de Calcul Intensif
  Aquitain) of the Universit\'e de Bordeaux and of the Universit\'e
  de Pau et des Pays de l'Adour. 
\end{acknowledgments}

\appendix

%
%
%
%

\section{One-pion exchange potential}

In momentum space the one-pion exchange (nonrelativistic) potential
takes the form
\begin{subequations}
\eq{
  \elmx{\vect k'}{\hat{V}^{(0)}_{1\pi}}{\vect k} =
  W_{T,1\pi}(q) \, (\vectop{\sigma}_1 \cdot \mathbf q) 
(\vectop{\sigma}_2 \cdot \mathbf q) \, (\vectop{\tau}_1 \cdot
\vectop{\tau}_2)
\label{V1pi_LO_repeated}
}
where $\vect k$, $\vect k'$ are incoming and outgoing relative momenta,
$\vectop{\sigma}_i$ and $\vectop{\tau}_i$
are Pauli spin and isospin operators, and the form factor $W_{T,1\pi}(q)$ reads
\eq{
\label{WT1pi_repeated}
W_{T,1\pi}(q) = 
-\Big(\dfrac{\gA}{2f_{\pi}}\Big)^2 \, 
\dfrac{(\hbar c)^3}{q^2 + \Lambda_{\pi}^2} \,.
}
\end{subequations}
This potential depends only on the momentum transfer $\vect q = \vect
k' - \vect k$ owing to its local character. We introduce the notation
$\Lambda_{\pi} = \dfrac{m_{\pi}c^2}{\hbar c}$ for the inverse of the
reduced Compton wavelength of the pion. Here $m_{\pi}$ represents the
mean pion mass. In coordinate space the one-pion exchange potential is
given by
\begin{subequations}
\eq{
  \elmx{\vect r'}{\hat{V}_{\rm LO}}{\vect r} =
  \delta(\vect r' - \vect r) V_{1\pi}(\vect r)
}
where
\begin{align}
  V_{1\pi}(\vect r) = & \frac{(2\pi)^3}{N_k} \, \frac{(m_{\pi}c^2)^3}{12 \pi} \,
  \Big(\frac{\gA}{2f_{\pi}}\Big)^2 \,
  (\vectop{\tau}_1 \cdot \vectop{\tau}_2) \,
  \Big[T_{1\pi}(\Lambda_{\pi}r) \, \hat S_{12} \nonumber \\
    & +
    \Big(Y_{1\pi}(\Lambda_{\pi}r) - \frac{4\pi}{\Lambda_{\pi}^3} \,
    \delta(\vect r)\Big) \, \vectop{\sigma}_1 \cdot \vectop{\sigma}_2
    \Big]\,,
\end{align}
\end{subequations}
The constant $N_k$ is a plane-wave normalization factor defined by
\begin{subequations}
\eq{
\overlap{\vect k'}{\vect k} = N_k \, \delta(\vect k' - \vect k) \,,
}
hence the following expression of the plane wave in coordinate space
\eq{
  \overlap{\vect r}{\vect k} = \sqrt{\frac{N_k}{(2\pi)^3}} \,
  e^{i\vect k \cdot \vect r}\,.
  }
\end{subequations}
The usual tensor operator $\hat S_{12}$ reads
\eq{
  \hat S_{12} = \frac 3{r^2} \, \tensorcoupling{\vect r}{\vect r}{2}
  \cdot \tensorcoupling{\vectop{\sigma}_1}{\vectop{\sigma}_2}{2}
  = 3 \, (\vectop{\sigma}_1 \cdot \hat r)
  (\vectop{\sigma}_2 \cdot \hat r) - \vectop{\sigma}_1 \cdot
  \vectop{\sigma}_2
}
where $\hat r = \vect r/r$ is unit vector and
$\tensorcoupling{A}{B}{k}$ denotes the irreducible tensor
product of rank $k$ of spherical tensors $A$ and $B$ according
to the notation and definition of Varshalovich et
al.~\cite{Varshalovich88}. The functions $Y_{1\pi}$ and $T_{1\pi}$ are
defined as
\begin{subequations}
  \begin{align}
    Y_{1\pi}(x) & = \frac{e^{-x}}x \\
    T_{1\pi}(x) & = Y_{1\pi}(x) \, \Big(1 + \frac 3 x + \frac 3 {x^2}\Big)\,.
  \end{align}
\end{subequations}

%
%
%
%

\section{Low-energy constants}

\paragraph{$^1S_0$ channel.}

\begin{figure}[h]
  \includegraphics[width=0.48\textwidth]{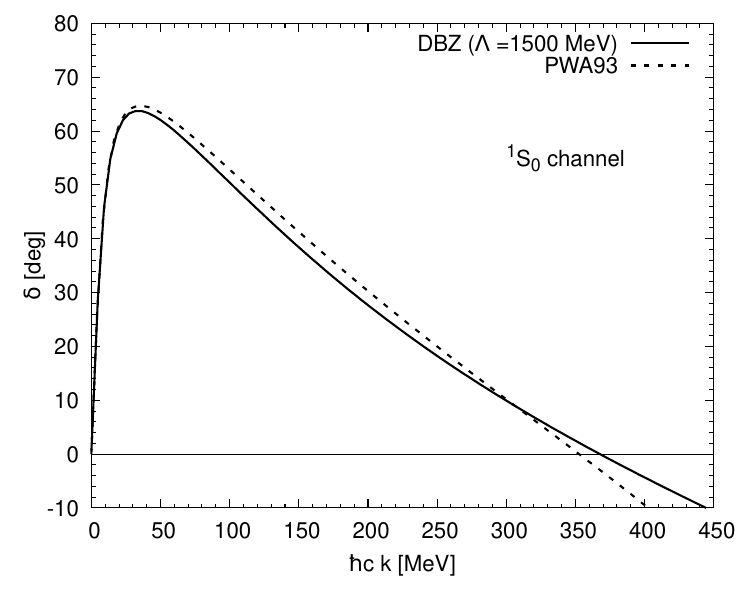}
  \caption{Phaseshift in the $^1S_0$ channel with the DBZ counter-term
    and the LECs values of Table~\ref{tab_LEC_1S0} for $\Lambda=1500$~MeV
  (solid line), compared with the empirical one from the Nijmegen
  multi-energy $np$ analysis~\cite{PWA93} (PWA93, dashed line).
\label{fig_best_phaseshift_1S0}}
\end{figure}

The three LECs $\Delta_1$, $\Delta_2$, $\gamma$ in Eq.~(\ref{eq_DBZ})
are fitted to reproduce the effective-range expansion (scattering
length $a_0= -23.75$~fm and effective-range parameter $r_0=2.77$~fm
as used in Ref.~\cite{Song17_PRC96}) and
the vanishing of the phase shift at $\hbar c \, k_0 \approx
350$~MeV as used in
Ref.~\cite{Sanchez20_PRC102}. The values of physical constants
retained in our adjustment of LECs are $\hbar c = 197.327$~$\rm MeV
\cdot fm$, $2\mu c^2 = 938.918$~MeV, $g_A = 1.26$, $f_{\pi} = 92.4$~MeV
and $m_{\pi}c^2 = 138.03$~MeV. Table~\ref{tab_LEC_1S0} shows the
resulting values for several values of the momentum cutoff $\Lambda$,
whereas figure~\ref{fig_best_phaseshift_1S0} shows the resulting
phaseshift for $\Lambda=1500$~MeV and illustrates in particular the
sucessful fit of $k_0$. 
\begin{table}[h]
  \centering
  \caption{Low-energy constants $\Delta_1$, $\Delta_2$ and $\gamma$ (in MeV)
    for various values of $\Lambda$ in Eq.~(\ref{eq_freg}).
    \label{tab_LEC_1S0}}
    \begin{tabular}{cccc}
      \hline \hline
      $\Lambda$ (MeV) & $\Delta_1$ (MeV) & $\Delta_2$ (MeV) & $\gamma$ (MeV) \\
      \hline
      500 & $-89.4$ & 473.0 & 312.0 \\
      1000 & $-81.2$ & 282.8 & 274.4 \\
      1500 & $-94.3$ & 319.1 & 243.1 \\
      \hline \hline
    \end{tabular}
\end{table}

For completeness we
also give the partial-wave matrix elements of the one-pion exchange
potential (including the regularization function) in $^1S_0$ and
$^3S_1$ channels
\begin{align}
  V_{1\pi}^{(^1S_0)}(k',k) = & V_{1\pi}^{(^3S_1)}(k',k) \nonumber \\
  = & \frac{4\pi N'_k}{N_k} \,
  \Big(\frac{g_A\mpi c^2}{2f_{\pi}}\Big)^2 \,
  \frac{\hbar c}{\Lambda_{\pi}^2} \,\bigg[1 -
    \frac{\Lambda_{\pi}^2}{4k'k} \times \nonumber \\
  & \ln \bigg(\frac{(k'+k)^2+\Lambda_{\pi}^{\,2}}{ 
      (k'-k)^2+\Lambda_{\pi}^{\,2}}\bigg)\bigg] \, f_{\rm reg}(k',k) \,.
  \label{eq_V1pi_1S0_k}
\end{align}

\paragraph{$^3S_1$ channel.}

In this work we choose to write regularized partial-wave matrix elements
of $\hat V_{\rm ct}$ in the $^3S_1$ channel as
\begin{align}
  V_{\rm ct}^{(^3\!S_1)}(k',k) & = \frac{4\pi N'_k}{N_k} \,
  C_{^3\!S_1} \, \frac{\hbar c}{\Lambda_{\pi}^2} \times f_{\rm reg}(k',k) \,.
\end{align}
The LEC $C_{^3\!S_1}$ is dimensionless and adjusted to the $np$
scattering length ($a^{(^3S_1)} = 5.42$~fm as used in Ref.~\cite{Nogga05_PRC72})
as a function of the regularization cutoff $\Lambda$ by solving
the Lippmann--Schwinger equation. The coupling to the $D$ wave
is taken into account. 

For the same values of $\Lambda$ as above, we find the LEC values reported
in table~\ref{tab_LEC_3S1_WPC}. We also provide the predicted
effective-range parameter $r_0$, reasonably independent of
$\Lambda$ and close to its value 1.75~fm deduced, e.g, from the Nijmegen
partial-wave analysis~\cite{PWA93}.
\begin{table}[h]
  \caption{Values of the low-energy constant in the $^3S_1$
    partial wave for various values of $\Lambda$ in
    Eq.~(\ref{eq_freg}) fitted to the scattering length and predicted
    effective-range parameter $r_0$. \label{tab_LEC_3S1_WPC}}
    \centering
    \begin{tabular}{cccc}
      \hline \hline
      $\Lambda$ (MeV) & $C_{^3S_1}$ & $C_{^3S_1}
      \dfrac{\hbar c}{\Lambda_{\pi}^2}$ ($\rm MeV \cdot fm^3$) & $r_0$ (fm)\\
      \hline
      500 & $-1.19815$ & $-483.20$ & 1.50 \\
      1000 & $+8.2743$ & $+3336.90$ & 1.65 \\
      1500 & $-2.36022$ & $-951.84$ & 1.63 \\
      \hline \hline
    \end{tabular}
\end{table}
\begin{table*}[t]
  \centering
  \caption{Low-energy constants in the $^3P_0$, $^3P_2$ and $^3D_2$
    channels for several values of $\Lambda$ in Eq.~(\ref{eq_freg}).
    \label{tab_LEC_3P0_3P2_3D2}}
    \begin{tabular}{c|cc|cc|cc}
      \hline \hline
      $\Lambda$ (MeV) &
      $C_{^3P_0}(\times 10^{-4})$ & $C_{^3P_0}
      \dfrac{\hbar c}{\Lambda_{\pi}^4}$ ($\rm MeV \cdot fm^5$) &
      $C_{^3P_2}(\times 10^{-4})$ & $C_{^3P_2}
      \dfrac{\hbar c}{\Lambda_{\pi}^4}$ ($\rm MeV \cdot fm^5$) &
      $C_{^3D_2}(\times 10^{-4})$ & $C_{^3D_2}
      \dfrac{\hbar c}{\Lambda_{\pi}^6}$ ($\rm MeV \cdot fm^7$) \\
      \hline
      500 & 2315.915 & 190.88 & $-732.28$ & $-60.36$ & $-31.643$ & $-5.33$ \\
      1000 & $-522.851$ & $-43.09$ & $-244.57$ & $-20.16$ & $+8.353$ & 1.41 \\
      1500 & $-3.189$ & $-0.26$ & $-88.942$ & $-7.33$ & $-3.214$ & $-0.54$\\
      \hline \hline
    \end{tabular}
\end{table*}
As expected from the singular and attractive character of
the tensor potential in the coupled partial waves, and as
already observed by Nogga and collaborators~\cite{Nogga05_PRC72},
the LEC in the $^3\!S_1$ channel exhibits a limit-cycle-like
behavior in its variation with~$\Lambda$. In particular these authors,
who use the regularization function of Eq.~(\ref{eq_freg}),
find a vertical asymptote for $\Lambda \approx 1250$~MeV. This explains
the large positive value of $C_{^3S_1}$ for $\Lambda = 1000$~MeV and
its large negative value for $\Lambda = 1500$~MeV.

\paragraph{$^3P_0$, $^3P_2$ and $^3D_2$ channels.}

The counter terms in these channels are written in the form
\begin{align}
\label{renormalized_3P0}
V_{\rm ct}^{(^3P_0)}(k',k) & = 4\pi \, \frac{N'_k}{N_k} \,
C_{^3P_0}\, \frac{\hbar c}{\Lambda_{\pi}^2} \,
\frac{k' k}{\Lambda_{\pi}^2} \\
\label{renormalized_3P2}
V_{\rm ct}^{(^3P_2)}(k',k) & = 4\pi \, \frac{N'_k}{N_k} \,
C_{^3P_2}\, \frac{\hbar c}{\Lambda_{\pi}^2} \,
\frac{k' k}{\Lambda_{\pi}^2} \\
\label{renormalized_3D2}
V_{\rm ct}^{(^3D_2)}(k',k) & = 4\pi \, \frac{N'_k}{N_k} \,
C_{^3D_2}\, \frac{\hbar c}{\Lambda_{\pi}^2} \,
\frac{(k' k)^2}{\Lambda_{\pi}^4} \,.
\end{align}
The dimensionless low-energy constants $C_{^3P_0}$ and
$C_{^3P_2}$ have been adjusted to the phase shifts at
$T_{lab}(k)=2(\hbar k)^2/(2\mu)=50$~MeV from the Nijmegen
partial-wave analysis~\cite{PWA93}, namely
$\delta^{(^3P_0)} = 10.7^{\circ}$ and $\delta^{(^3P_2)} = 5.9^{\circ}$,
whereas $C_{^3D_2}$
has been adjusted to the phase shift at
$T_{lab}(k)=2(\hbar k)^2/(2\mu)=100$~MeV, namely
$\delta^{(^3D_2)}(k) = 17.3^{\circ}$. The results of the fits are
displayed in Table~\ref{tab_LEC_3P0_3P2_3D2}.

%
%
%
%

\section{Confined plane-wave basis}

\subsection{Basis wave functions confined in a box}

In coordinate
space these confined plane waves take the form
\eq{
  \overlap{\vect r}{\varphi_{\alpha}} = \varphi_{\alpha}(\vect r) =
  \begin{cases}
    L^{-3/2}e^{i\vect k_{\alpha} \cdot \vect r} & \mbox{if $\vect r \in
      \Big[-\dfrac L 2;\dfrac L 2\Big]^3$} \\
    0 & \mbox{otherwise.}
  \end{cases}
}
where $L$ is the edge length and $\alpha$ is a triplet of indices
$(\alpha_x,\alpha_y,\alpha_z)$ specified by the orthogonality condition
\eq{
  \overlap{\varphi_{\alpha'}}{\varphi_{\alpha}} = \delta_{\alpha'\alpha} =
  \delta_{\alpha'_x\alpha_x}\delta_{\alpha'_y\alpha_y}\delta_{\alpha'_z\alpha_z} \,.
}
This condition is satisfied in particular for momentum vectors
$\vect k_{\alpha} = (k_{\alpha_x},k_{\alpha_y},k_{\alpha_z})$
(in the cartesian basis) such that $k_{\alpha_x} = \alpha_x \,
\dfrac{2\pi}L$ with either $\alpha_x \in \mathbb Z$ or
$\alpha_x - \dfrac 1 2 \in \mathbb Z$, and similarly for the other space
directions. Here we choose the latter case to obtain non vanishing
momentum values. We thus denote by $\ket{\varphi_{\alpha}\sigma\tau}$ the basis
states, which are infinite in number up to this point. 

It turns out that, in one dimension (for example the $x$ direction),
the wave function
\eq{
  \varphi_{\alpha_x}(x) =
  \begin{cases}
    L^{-1/2}e^{ik_{\alpha_x}x} & \mbox{if $x \in \Big[-\dfrac L
        2;\dfrac L 2\Big]$} \\
    0 & \mbox{otherwise} \,,
  \end{cases}
}
is the
variational basis of the Lagrange-mesh method associated with an 
equidistant mesh on the $x$ axis over the interval $\big[-\frac L 2;
+\frac L 2\big]$~\cite{Baye15_PhysRep565}. This finite mesh, called
the Lagrange-Fourier mesh, is made of the $N$ abscissa $x_m = m\,dx$
with the mesh step $dx = \dfrac L N$ and $m \in I_N$ where 
\eq{
  I_N = \Big\{-\dfrac{N-1}2, -\dfrac{N-1}2+1,
  ... \dfrac{N-1}2\Big\} \,.
}
The Lagrange functions $f_j(x)$, with $j \in
I_N$, associated with this mesh are such that $f_j(x_m) = \delta_{jm}$
and are orthogonal to each other at the Gauss-quadrature approximation
based on this mesh. The two sets of functions $\{f_j(x),j\in I_N\}$
and $\{\varphi_{\alpha_x}(x), \alpha_x\in I_N\}$ are related by a unitary
transformation (see section 2.2 of Ref.~\cite{Baye15_PhysRep565}). 
This offers a dual interpretation of the confined plane-wave basis and
its parameters $L$ and $N$ (or equivalently $L$ and $dx$ such that
$L/dx \in \mathbb N^*$). In all rigor the set of orthonormal states
$\{\ket{\varphi_{\alpha_x}},\alpha_x \in I_N\}$
is a Hilbert basis of the Hilbert space of square integrable functions
over the interval $\big[-\frac L 2; +\frac L 2\big]$ only in the limit
$N \to \infty$. 

Because $(I_N,N\in \mathbb N^*,\subset)$ is an increasing sequence for
inclusion, the parameter $N$ (or $dx$) for a fixed edge length $L$ is
a variational parameter for Hartree--Fock calculations. Indeed 
increasing $N$ brings additional states in the single-particle basis
and thus always yield ground-state solutions lower in energy. 
In contrast increasing $L$ does not necessarily produce lower-energy
solutions, but approximate solutions expectedly closer to the exact
ones which correspond to an infinite box size.

\subsection{Symmetry-adapted basis}

The set of
$\ket{\varphi_{\alpha}\sigma}$ for a fixed isospin projection, 
corresponding to either neutron or proton single-particle states, 
is a basis of reducible co-representation of the full octahedral
double group with time-reversal symmetry, denoted by $O_{\rm
  2h}^{\rm DT}$. We build a symmetry-adapted basis by a subsequent unitary
transformation so that the resulting orthonormal set of states is a
basis of reducible co-representation of a subgroup $G$ of $O_{\rm 2h}^{\rm
DT}$. We choose here the subgroup $G = \mathrm{Gr}\{\hat{\Pi}, \hat
R_z,\hat R_y^T \}$ generated by the parity operator $\hat{\Pi}$, the
$z$-signature operator $\hat R_z$ and the anitunitary operator
$\hat R_y^T = \hat{\mathcal T}\hat R_y$ where $\hat{\mathcal T}$ is the time-reversal
operator. Even if we consider, in this work, time-reversal invariant
solutions only, it is more advantageous not to add $\hat{\mathcal T}$ to the
above symmetry group $G$, which would yield the
full dihedral double group with time-reversal symmetry
$D_{2h}^{\rm DT}$. Indeed the unitary subgroup $\mathrm{Gr}\{\hat\Pi,\hat
R_z\}$ of $G$ is abelian and yields two quantum numbers
(intrinsic parity and $z$-signature), whereas the unitary subgroup
$\mathrm{Gr}\{\hat\Pi,\hat R_z,\hat R_y\}$ of $D_{2h}^{\rm DT}$ is non
abelian and yields only one quantum number (intrinsic
parity). Moreover in both cases one can reduce the set of discretized
momenta to one eighth of the full three-dimensional mesh in order to
generate a basis of corepresentation of $O^{\rm DT}_{2h}$.

The symmetry-adapted basis in the present case is obtained by a proper
unitary transformation of the above defined confined plane-wave
basis. It can be constructed by use of projection operators
\begin{subequations}
\begin{align}
\hat P(p) & = \dfrac 1 2 \Big(\mathbbm 1 + \dfrac{\hat\Pi}{p}\Big) \\
\hat P(r_z) & = \dfrac 1 2 \Big(\mathbbm 1 + \dfrac{\hat
  R_z}{r_z}\Big)
\end{align}
\end{subequations}
where $p=\pm 1$ and $r_z=\pm i$ are the intrinsic-parity and
$z$-signature quantum numbers. Moreover, defining the operator
$\hat Q_c = \dfrac 1 2 (\mathbbm 1 + c\hat R_y^T)$,
with $c=\pm 1$, one can show that for fixed $(p,r_z)$, $\alpha =
(\alpha_x,\alpha_y,\alpha_z)$, with $\alpha_m-\dfrac 1 2 \in \mathbbm N$
($m=x,y,z$) and $\sigma=\pm \dfrac 1 2$, the two states 
\eq{
    \ket{pr_z(c\alpha\sigma)} = \sqrt 8 \: 
    \hat P(p) \, \hat P(r_z) \, \hat Q_c \, \ket{\varphi_\alpha \sigma}\,, 
\quad c = \pm 1\,,
\label{symmetry_adapted_state}
}
form two bases of equivalent irreducible corepresentations of $G$
of dimension 1. Note that here we have chosen to work with $\alpha$
being a half-integer triplet, that is even $N$ values, therefore all
$\alpha_m$ indices are different from 0.

\subsection{Linear-momentum property of the confined 
plane-wave basis}

We first write the matrix element of a two-body potential $\hat V$ between
confined plane-wave states through the coordinate representation of
these states as
\begin{widetext}
\eq{
\elmx{\varphi_{\alpha'_1}\varphi_{\alpha'_2}}{\hat
    V}{\varphi_{\alpha_1}\varphi_{\alpha_2}} = L^{-6} 
  \int_{\mathcal C_{L/2}}d^3\vect r'_1
  \int_{\mathcal C_{L/2}}d^3\vect r'_2
  \int_{\mathcal C_{L/2}}d^3\vect r_1
  \int_{\mathcal C_{L/2}}d^3\vect r_2 \: e^{i(\vect k_{\alpha_1}\cdot \vect r_1
    + \vect k_{\alpha_2}\cdot \vect r_2
    - \vect k_{\alpha'_1}\cdot \vect r'_1
    - \vect k_{\alpha'_2}\cdot \vect r'_2)} \elmx{\vect r'_1\vect
    r'_2}{\hat V}{\vect r_1\vect r_2} \,,
  \label{eq_v_1}
}
\end{widetext}
where $\mathcal C_{L/2} = \big[-\frac L2 ; +\frac L2\big]^3$.
Owing to Galilean and translation invariances of $\hat V$,
the matrix element $\elmx{\vect r'_1\vect r'_2}{\hat V}{\vect r_1\vect
  r_2}$ is of the form
\eq{
  \elmx{\vect r'_1\vect r'_2}{\hat V}{\vect r_1\vect r_2} = 
  \delta(\vect R' - \vect R)\elmx{\vect r'}{\hat V}{\vect r}
}
where $\vect R = \frac 1 2 \,(\vect r_1 + \vect r_2)$ is the center-of-mass
position vector and $\vect r = \vect r_1 - \vect r_2$ (similarly for
$\vect r'$) is the relative position vector. Using these substitutions
in the integrals of Eq.~(\ref{eq_v_1}) we obtain
\begin{align}
\elmx{\varphi_{\alpha'_1}\varphi_{\alpha'_2}}{\hat
    V}{\varphi_{\alpha_1}\varphi_{\alpha_2}} = & 
L^{-6}\int_{\mathcal C_L}d^3\vect r'\int_{\mathcal C_L}d^3\vect r \:
  e^{i(\vect k_{\alpha} \cdot \vect r - \vect k_{\alpha'} \cdot \vect
    r')}\times\nonumber \\
&  \elmx{\vect r'}{\hat V}{\vect r}
\int_{\mathcal D' \cap \mathcal D} d^3\vect R \:
  e^{i(\vect K_{\alpha'} - \vect K_{\alpha}) \cdot \vect R}
   \label{eq_v_2} 
\end{align}
where the domains of integration $\mathcal C_L$, $\mathcal D$
and $\mathcal D'$ are defined by
\begin{subequations}
  \begin{align}
  \mathcal C_L & = \Big[-\frac L2 ; +\frac L2\Big]^3 \\
  \mathcal D & = \mathcal D_x \times \mathcal D_y \times \mathcal D_z \\
  \mathcal D' & = \mathcal D_{x'} \times \mathcal D_{y'} \times
  \mathcal D_{z'}
  \end{align}
\end{subequations}
with, for instance,
\eq{
\mathcal D_x = \Big[-\frac L 2+\frac{|x|}2;\frac L 2-\frac{|x|}2\Big] \,.
}
Moreover in Eq.~(\ref{eq_v_2}) we have
introduced the incoming relative momentum $\vect k_{\alpha}$ and
incoming total momentum~$\vect K_{\alpha}$ defined by
\begin{align}
  \label{eq_relative_momentum}
\vect k_{\alpha} & = \frac 1 2 \, (\vect k_{\alpha_1} - \vect k_{\alpha_2}) \\
\vect K_{\alpha} & = \vect k_{\alpha_1} + \vect k_{\alpha_2} \,,
\end{align}
and similarly for outgoing momenta. Owing to the short range of the
two-nucleon strong interaction $\hat V$, we can approximate the domains
$\mathcal D$ and $\mathcal D'$ with $\mathbb R^3$ when the box size
$L$ is much larger than the range of $\hat V$. In this approximation the
domain of integration over $\vect R$ then becomes $\mathbb R^3$ and
the integrals over $\vect R$ on the one hand, and over $\vect r'$,
$\vect r$ on the other hand, can be factorized. We end up with
\begin{align}
  \elmx{\varphi_{\alpha'_1}\varphi_{\alpha'_2}}{\hat
    V}{\varphi_{\alpha_1} \varphi_{\alpha_2}} \approx &\, 
  \delta_{\alpha'_1+\alpha'_2 \, \alpha_1+\alpha_2} \, 
  \frac{1}{N_k} \,
  \bigg(\frac{2\pi}{L}\bigg)^3 \times \nonumber \\
  & \elmx{\vect k_{\alpha'}}{\hat
    V}{\vect k_{\alpha}} \,,\label{eq_v_4}
\end{align}
where $\alpha_1+\alpha_2$ is the triplet of integers
$(\alpha_{1x}+\alpha_{2x},\alpha_{1y}+\alpha_{2y},\alpha_{1z}+\alpha_{2z})$,
and the relative-momentum multi-index $\alpha$ reads
$$
\alpha = \frac 1 2 \, (\alpha_1 - \alpha_2) = 
\Big(\frac{\alpha_{1x} -
  \alpha_{2x}}2,\frac{\alpha_{1y}-\alpha_{2y}}2,
\frac{\alpha_{1z}-\alpha_{2z}}2\Big)\,.
$$
Therefore in the confined
plane-wave basis, the two-body matrix elements are simply proportional
to the momentum representation of the interaction. This is in
constrast with other bases, such as the momentum partial-wave or
harmonic-oscillator basis, where it is neccessary to perform a 
transformation from the laboratory frame to the center-of-mass frame
using vector brackets~\cite{Wong72_NPA183} or Moshinsky
coefficients~\cite{Moshinsky59_NP13}, respectively.

In addition to a simplified calculation, the confined
plane-wave basis offers an economical representation of the two-body
matrix elements of $\hat V$. Indeed the set of distinct relative
momenta in one dimension $k_{\alpha_x} = \frac 1 2 \, (k_{\alpha_{1x}}
- k_{\alpha_{2x}})$ generated by a set of equidistant single-particle
momenta $k_{\alpha_{1x}}$ and $k_{\alpha_{2x}}$ (see subsection II.A)
is $\{\frac{\pi}L \, \alpha_x, \, -N+1\leqslant \alpha_x\leqslant
N-1\}$ (by unit step), the cardinal of which is $2N-1$ instead of
$N^2$ for a non-equidistant momentum set. This entails a considerable
gain of memory to store the (antisymmetrized) two-body matrix elements
of $\hat V$ in the confined plane-wave basis.

\input{LO_HF_16Apr2023_PRC.bbl}

\end{document}

%% file: LO_HF_16Apr2023_PRC.bbl
%

%% file: LO_HF_16Apr2023_PRC_resubmitted.bbl
\begin{thebibliography}{52}%
\makeatletter
\providecommand \@ifxundefined [1]{%
 \@ifx{#1\undefined}
}%
\providecommand \@ifnum [1]{%
 \ifnum #1\expandafter \@firstoftwo
 \else \expandafter \@secondoftwo
 \fi
}%
\providecommand \@ifx [1]{%
 \ifx #1\expandafter \@firstoftwo
 \else \expandafter \@secondoftwo
 \fi
}%
\providecommand \natexlab [1]{#1}%
\providecommand \enquote  [1]{``#1''}%
\providecommand \bibnamefont  [1]{#1}%
\providecommand \bibfnamefont [1]{#1}%
\providecommand \citenamefont [1]{#1}%
\providecommand \href@noop [0]{\@secondoftwo}%
\providecommand \href [0]{\begingroup \@sanitize@url \@href}%
\providecommand \@href[1]{\@@startlink{#1}\@@href}%
\providecommand \@@href[1]{\endgroup#1\@@endlink}%
\providecommand \@sanitize@url [0]{\catcode `\\12\catcode `\$12\catcode
  `\&12\catcode `\#12\catcode `\^12\catcode `\_12\catcode `\%12\relax}%
\providecommand \@@startlink[1]{}%
\providecommand \@@endlink[0]{}%
\providecommand \url  [0]{\begingroup\@sanitize@url \@url }%
\providecommand \@url [1]{\endgroup\@href {#1}{\urlprefix }}%
\providecommand \urlprefix  [0]{URL }%
\providecommand \Eprint [0]{\href }%
\providecommand \doibase [0]{http://dx.doi.org/}%
\providecommand \selectlanguage [0]{\@gobble}%
\providecommand \bibinfo  [0]{\@secondoftwo}%
\providecommand \bibfield  [0]{\@secondoftwo}%
\providecommand \translation [1]{[#1]}%
\providecommand \BibitemOpen [0]{}%
\providecommand \bibitemStop [0]{}%
\providecommand \bibitemNoStop [0]{.\EOS\space}%
\providecommand \EOS [0]{\spacefactor3000\relax}%
\providecommand \BibitemShut  [1]{\csname bibitem#1\endcsname}%
\let\auto@bib@innerbib\@empty
\bibitem [{\citenamefont {Weinberg}(1990)}]{Weinberg90_PLB251}%
  \BibitemOpen
  \bibfield  {author} {\bibinfo {author} {\bibfnamefont {S.}~\bibnamefont
  {Weinberg}},\ }\href@noop {} {\bibfield  {journal} {\bibinfo  {journal}
  {Phys. Lett. B}\ }\textbf {\bibinfo {volume} {251}},\ \bibinfo {pages} {288}
  (\bibinfo {year} {1990})}\BibitemShut {NoStop}%
\bibitem [{\citenamefont {Weinberg}(1991)}]{Weinberg91_NPB363}%
  \BibitemOpen
  \bibfield  {author} {\bibinfo {author} {\bibfnamefont {S.}~\bibnamefont
  {Weinberg}},\ }\href@noop {} {\bibfield  {journal} {\bibinfo  {journal}
  {Nucl. Phys. B}\ }\textbf {\bibinfo {volume} {363}},\ \bibinfo {pages} {3}
  (\bibinfo {year} {1991})}\BibitemShut {NoStop}%
\bibitem [{\citenamefont {Machleidt}\ and\ \citenamefont
  {Entem}(2011)}]{Machleidt11_PhysRep503}%
  \BibitemOpen
  \bibfield  {author} {\bibinfo {author} {\bibfnamefont {R.}~\bibnamefont
  {Machleidt}}\ and\ \bibinfo {author} {\bibfnamefont {D.~R.}\ \bibnamefont
  {Entem}},\ }\href@noop {} {\bibfield  {journal} {\bibinfo  {journal} {Phys.
  Rep.}\ }\textbf {\bibinfo {volume} {503}},\ \bibinfo {pages} {1} (\bibinfo
  {year} {2011})}\BibitemShut {NoStop}%
\bibitem [{\citenamefont {Gysbers}\ \emph {et~al.}(2019)\citenamefont
  {Gysbers}, \citenamefont {Hagen}, \citenamefont {Holt}, \citenamefont
  {Jansen}, \citenamefont {Morris}, \citenamefont {Navr\'atil}, \citenamefont
  {Papenbrock}, \citenamefont {Quaglioni}, \citenamefont {Schwenk},
  \citenamefont {Stroberg},\ and\ \citenamefont {Wendt}}]{Gysbers19_Nature15}%
  \BibitemOpen
  \bibfield  {author} {\bibinfo {author} {\bibfnamefont {P.}~\bibnamefont
  {Gysbers}}, \bibinfo {author} {\bibfnamefont {G.}~\bibnamefont {Hagen}},
  \bibinfo {author} {\bibfnamefont {J.~D.}\ \bibnamefont {Holt}}, \bibinfo
  {author} {\bibfnamefont {G.~R.}\ \bibnamefont {Jansen}}, \bibinfo {author}
  {\bibfnamefont {T.~D.}\ \bibnamefont {Morris}}, \bibinfo {author}
  {\bibfnamefont {P.}~\bibnamefont {Navr\'atil}}, \bibinfo {author}
  {\bibfnamefont {T.}~\bibnamefont {Papenbrock}}, \bibinfo {author}
  {\bibfnamefont {S.}~\bibnamefont {Quaglioni}}, \bibinfo {author}
  {\bibfnamefont {A.}~\bibnamefont {Schwenk}}, \bibinfo {author} {\bibfnamefont
  {S.~R.}\ \bibnamefont {Stroberg}}, \ and\ \bibinfo {author} {\bibfnamefont
  {K.~A.}\ \bibnamefont {Wendt}},\ }\href@noop {} {\bibfield  {journal}
  {\bibinfo  {journal} {Nature Physics}\ }\textbf {\bibinfo {volume} {15}},\
  \bibinfo {pages} {428} (\bibinfo {year} {2019})}\BibitemShut {NoStop}%
\bibitem [{\citenamefont {Som{\`a}}\ \emph {et~al.}(2020)\citenamefont
  {Som{\`a}}, \citenamefont {Navr{\'a}til}, \citenamefont {Raimondi},
  \citenamefont {Barbieri},\ and\ \citenamefont {Duguet}}]{Soma20_PRC102}%
  \BibitemOpen
  \bibfield  {author} {\bibinfo {author} {\bibfnamefont {V.}~\bibnamefont
  {Som{\`a}}}, \bibinfo {author} {\bibfnamefont {P.}~\bibnamefont
  {Navr{\'a}til}}, \bibinfo {author} {\bibfnamefont {F.}~\bibnamefont
  {Raimondi}}, \bibinfo {author} {\bibfnamefont {C.}~\bibnamefont {Barbieri}},
  \ and\ \bibinfo {author} {\bibfnamefont {T.}~\bibnamefont {Duguet}},\
  }\href@noop {} {\bibfield  {journal} {\bibinfo  {journal} {Phys. Rev. C}\
  }\textbf {\bibinfo {volume} {102}},\ \bibinfo {pages} {014318} (\bibinfo
  {year} {2020})}\BibitemShut {NoStop}%
\bibitem [{\citenamefont {Hergert}(2020)}]{Hergert20_FrontPhys8}%
  \BibitemOpen
  \bibfield  {author} {\bibinfo {author} {\bibfnamefont {H.}~\bibnamefont
  {Hergert}},\ }\href@noop {} {\bibfield  {journal} {\bibinfo  {journal}
  {Front. Phys.}\ }\textbf {\bibinfo {volume} {8}},\ \bibinfo {pages} {379}
  (\bibinfo {year} {2020})}\BibitemShut {NoStop}%
\bibitem [{\citenamefont {Maris}\ \emph {et~al.}(2020)\citenamefont {Maris},
  \citenamefont {Epelbaum}, \citenamefont {Furnstahl}, \citenamefont {Golak},
  \citenamefont {Hebeler}, \citenamefont {H{\"u}ther}, \citenamefont {Kamada},
  \citenamefont {Krebs}, \citenamefont {Mei{\ss}ner}, \citenamefont {Melendez},
  \citenamefont {Nogga}, \citenamefont {Reinert}, \citenamefont {Roth},
  \citenamefont {Skibi{\'n}ski}, \citenamefont {Soloviov}, \citenamefont
  {Topolnicki}, \citenamefont {Vary}, \citenamefont {Volkotrub}, \citenamefont
  {Wita{\l}a},\ and\ \citenamefont {Wolfgruber}}]{Maris20_arXiv}%
  \BibitemOpen
  \bibfield  {author} {\bibinfo {author} {\bibfnamefont {P.}~\bibnamefont
  {Maris}}, \bibinfo {author} {\bibfnamefont {E.}~\bibnamefont {Epelbaum}},
  \bibinfo {author} {\bibfnamefont {R.~J.}\ \bibnamefont {Furnstahl}}, \bibinfo
  {author} {\bibfnamefont {J.}~\bibnamefont {Golak}}, \bibinfo {author}
  {\bibfnamefont {K.}~\bibnamefont {Hebeler}}, \bibinfo {author} {\bibfnamefont
  {T.}~\bibnamefont {H{\"u}ther}}, \bibinfo {author} {\bibfnamefont
  {H.}~\bibnamefont {Kamada}}, \bibinfo {author} {\bibfnamefont
  {H.}~\bibnamefont {Krebs}}, \bibinfo {author} {\bibfnamefont {U.-G.}\
  \bibnamefont {Mei{\ss}ner}}, \bibinfo {author} {\bibfnamefont {J.~A.}\
  \bibnamefont {Melendez}}, \bibinfo {author} {\bibfnamefont {A.}~\bibnamefont
  {Nogga}}, \bibinfo {author} {\bibfnamefont {P.}~\bibnamefont {Reinert}},
  \bibinfo {author} {\bibfnamefont {R.}~\bibnamefont {Roth}}, \bibinfo {author}
  {\bibfnamefont {R.}~\bibnamefont {Skibi{\'n}ski}}, \bibinfo {author}
  {\bibfnamefont {V.}~\bibnamefont {Soloviov}}, \bibinfo {author}
  {\bibfnamefont {K.}~\bibnamefont {Topolnicki}}, \bibinfo {author}
  {\bibfnamefont {J.~P.}\ \bibnamefont {Vary}}, \bibinfo {author}
  {\bibfnamefont {Y.}~\bibnamefont {Volkotrub}}, \bibinfo {author}
  {\bibfnamefont {H.}~\bibnamefont {Wita{\l}a}}, \ and\ \bibinfo {author}
  {\bibfnamefont {T.}~\bibnamefont {Wolfgruber}},\ }\href@noop {} {\bibfield
  {journal} {\bibinfo  {journal} {arXiv:2012.12396v2}\ } (\bibinfo {year}
  {2020})}\BibitemShut {NoStop}%
\bibitem [{\citenamefont {Nogga}\ \emph {et~al.}(2005)\citenamefont {Nogga},
  \citenamefont {Timmermans},\ and\ \citenamefont {van Kolck}}]{Nogga05_PRC72}%
  \BibitemOpen
  \bibfield  {author} {\bibinfo {author} {\bibfnamefont {A.}~\bibnamefont
  {Nogga}}, \bibinfo {author} {\bibfnamefont {R.~G.~E.}\ \bibnamefont
  {Timmermans}}, \ and\ \bibinfo {author} {\bibfnamefont {U.}~\bibnamefont {van
  Kolck}},\ }\href@noop {} {\bibfield  {journal} {\bibinfo  {journal} {Phys.
  Rev. C}\ }\textbf {\bibinfo {volume} {72}},\ \bibinfo {pages} {054006}
  (\bibinfo {year} {2005})}\BibitemShut {NoStop}%
\bibitem [{\citenamefont {Machleidt}\ \emph {et~al.}(2010)\citenamefont
  {Machleidt}, \citenamefont {Liu}, \citenamefont {Entem},\ and\ \citenamefont
  {Arriola}}]{Machleidt10_PRC81}%
  \BibitemOpen
  \bibfield  {author} {\bibinfo {author} {\bibfnamefont {R.}~\bibnamefont
  {Machleidt}}, \bibinfo {author} {\bibfnamefont {P.}~\bibnamefont {Liu}},
  \bibinfo {author} {\bibfnamefont {D.~R.}\ \bibnamefont {Entem}}, \ and\
  \bibinfo {author} {\bibfnamefont {E.~R.}\ \bibnamefont {Arriola}},\
  }\href@noop {} {\bibfield  {journal} {\bibinfo  {journal} {Phys. Rev. C}\
  }\textbf {\bibinfo {volume} {81}},\ \bibinfo {pages} {024001} (\bibinfo
  {year} {2010})}\BibitemShut {NoStop}%
\bibitem [{\citenamefont {Song}\ \emph {et~al.}(2017)\citenamefont {Song},
  \citenamefont {Lazauskas},\ and\ \citenamefont {van Kolck}}]{Song17_PRC96}%
  \BibitemOpen
  \bibfield  {author} {\bibinfo {author} {\bibfnamefont {Y.-H.}\ \bibnamefont
  {Song}}, \bibinfo {author} {\bibfnamefont {R.}~\bibnamefont {Lazauskas}}, \
  and\ \bibinfo {author} {\bibfnamefont {U.}~\bibnamefont {van Kolck}},\
  }\href@noop {} {\bibfield  {journal} {\bibinfo  {journal} {Phys. Rev. C}\
  }\textbf {\bibinfo {volume} {96}},\ \bibinfo {pages} {024002} (\bibinfo
  {year} {2017})}\BibitemShut {NoStop}%
\bibitem [{\citenamefont {Song}\ \emph {et~al.}(2019)\citenamefont {Song},
  \citenamefont {Lazauskas},\ and\ \citenamefont {van Kolck}}]{Song19_PRC100}%
  \BibitemOpen
  \bibfield  {author} {\bibinfo {author} {\bibfnamefont {Y.-H.}\ \bibnamefont
  {Song}}, \bibinfo {author} {\bibfnamefont {R.}~\bibnamefont {Lazauskas}}, \
  and\ \bibinfo {author} {\bibfnamefont {U.}~\bibnamefont {van Kolck}},\
  }\href@noop {} {\bibfield  {journal} {\bibinfo  {journal} {Phys. Rev. C}\
  }\textbf {\bibinfo {volume} {100}},\ \bibinfo {pages} {019901(E)} (\bibinfo
  {year} {2019})}\BibitemShut {NoStop}%
\bibitem [{\citenamefont {Yang}\ \emph
  {et~al.}(2021{\natexlab{a}})\citenamefont {Yang}, \citenamefont
  {Ekstr{\"o}m}, \citenamefont {Forss\'en},\ and\ \citenamefont
  {Hagen}}]{Yang21_PRC103}%
  \BibitemOpen
  \bibfield  {author} {\bibinfo {author} {\bibfnamefont {C.-J.}\ \bibnamefont
  {Yang}}, \bibinfo {author} {\bibfnamefont {A.}~\bibnamefont {Ekstr{\"o}m}},
  \bibinfo {author} {\bibfnamefont {C.}~\bibnamefont {Forss\'en}}, \ and\
  \bibinfo {author} {\bibfnamefont {G.}~\bibnamefont {Hagen}},\ }\href@noop {}
  {\bibfield  {journal} {\bibinfo  {journal} {Phys. Rev. C}\ }\textbf {\bibinfo
  {volume} {103}},\ \bibinfo {pages} {054304} (\bibinfo {year}
  {2021}{\natexlab{a}})}\BibitemShut {NoStop}%
\bibitem [{\citenamefont {Yang}\ \emph
  {et~al.}(2021{\natexlab{b}})\citenamefont {Yang}, \citenamefont
  {Ekstr{\"o}m}, \citenamefont {Forss\'en}, \citenamefont {Rupak},\ and\
  \citenamefont {van Kolck}}]{Yang21_arXiv2109.13303v1}%
  \BibitemOpen
  \bibfield  {author} {\bibinfo {author} {\bibfnamefont {C.-J.}\ \bibnamefont
  {Yang}}, \bibinfo {author} {\bibfnamefont {A.}~\bibnamefont {Ekstr{\"o}m}},
  \bibinfo {author} {\bibfnamefont {C.}~\bibnamefont {Forss\'en}}, \bibinfo
  {author} {\bibfnamefont {G.~H.~G.}\ \bibnamefont {Rupak}}, \ and\ \bibinfo
  {author} {\bibfnamefont {U.}~\bibnamefont {van Kolck}},\ }\href@noop {}
  {\bibfield  {journal} {\bibinfo  {journal} {2109.13303v1}\ } (\bibinfo {year}
  {2021}{\natexlab{b}})}\BibitemShut {NoStop}%
\bibitem [{\citenamefont {Epelbaum}\ and\ \citenamefont
  {Mei{\ss}ner}(2013)}]{Epelbaum06_arXiv0609037v2}%
  \BibitemOpen
  \bibfield  {author} {\bibinfo {author} {\bibfnamefont {E.}~\bibnamefont
  {Epelbaum}}\ and\ \bibinfo {author} {\bibfnamefont {U.-G.}\ \bibnamefont
  {Mei{\ss}ner}},\ }\href@noop {} {\bibfield  {journal} {\bibinfo  {journal}
  {arXiv:0609037v2}\ } (\bibinfo {year} {2013})}\BibitemShut {NoStop}%
\bibitem [{\citenamefont {Long}(2016)}]{Long16_EPJA25}%
  \BibitemOpen
  \bibfield  {author} {\bibinfo {author} {\bibfnamefont {B.}~\bibnamefont
  {Long}},\ }\href@noop {} {\bibfield  {journal} {\bibinfo  {journal} {Eur.
  Phys. J. E}\ }\textbf {\bibinfo {volume} {25}},\ \bibinfo {pages} {1641006}
  (\bibinfo {year} {2016})}\BibitemShut {NoStop}%
\bibitem [{\citenamefont {Epelbaum}\ \emph {et~al.}(2018)\citenamefont
  {Epelbaum}, \citenamefont {Gasparyan}, \citenamefont {Gegelia},\ and\
  \citenamefont {Mei{\ss}ner}}]{Epelbaum18_EPJA54}%
  \BibitemOpen
  \bibfield  {author} {\bibinfo {author} {\bibfnamefont {E.}~\bibnamefont
  {Epelbaum}}, \bibinfo {author} {\bibfnamefont {A.~M.}\ \bibnamefont
  {Gasparyan}}, \bibinfo {author} {\bibfnamefont {J.}~\bibnamefont {Gegelia}},
  \ and\ \bibinfo {author} {\bibfnamefont {U.-G.}\ \bibnamefont
  {Mei{\ss}ner}},\ }\href@noop {} {\bibfield  {journal} {\bibinfo  {journal}
  {Eur. Phys. J. A}\ }\textbf {\bibinfo {volume} {54}},\ \bibinfo {pages} {186}
  (\bibinfo {year} {2018})}\BibitemShut {NoStop}%
\bibitem [{\citenamefont {Valderrama}(2019)}]{Valderrama19_EPJA55}%
  \BibitemOpen
  \bibfield  {author} {\bibinfo {author} {\bibfnamefont {M.~P.}\ \bibnamefont
  {Valderrama}},\ }\href@noop {} {\bibfield  {journal} {\bibinfo  {journal}
  {Eur. Phys. J. A}\ }\textbf {\bibinfo {volume} {55}},\ \bibinfo {pages} {55}
  (\bibinfo {year} {2019})}\BibitemShut {NoStop}%
\bibitem [{\citenamefont {Epelbaum}\ \emph {et~al.}(2019)\citenamefont
  {Epelbaum}, \citenamefont {Gasparyan}, \citenamefont {Gegelia},\ and\
  \citenamefont {Mei{\ss}ner}}]{Epelbaum19_EPJA55}%
  \BibitemOpen
  \bibfield  {author} {\bibinfo {author} {\bibfnamefont {E.}~\bibnamefont
  {Epelbaum}}, \bibinfo {author} {\bibfnamefont {A.~M.}\ \bibnamefont
  {Gasparyan}}, \bibinfo {author} {\bibfnamefont {J.}~\bibnamefont {Gegelia}},
  \ and\ \bibinfo {author} {\bibfnamefont {U.-G.}\ \bibnamefont
  {Mei{\ss}ner}},\ }\href@noop {} {\bibfield  {journal} {\bibinfo  {journal}
  {Eur. Phys. J. A}\ }\textbf {\bibinfo {volume} {55}},\ \bibinfo {pages} {56}
  (\bibinfo {year} {2019})}\BibitemShut {NoStop}%
\bibitem [{\citenamefont {van Kolck}(2020)}]{vanKolck20_FrontPhys8}%
  \BibitemOpen
  \bibfield  {author} {\bibinfo {author} {\bibfnamefont {U.}~\bibnamefont {van
  Kolck}},\ }\href@noop {} {\bibfield  {journal} {\bibinfo  {journal} {Front.
  Phys.}\ }\textbf {\bibinfo {volume} {8}},\ \bibinfo {pages} {79} (\bibinfo
  {year} {2020})}\BibitemShut {NoStop}%
\bibitem [{\citenamefont {Slater}(1951)}]{Slater51_PR81}%
  \BibitemOpen
  \bibfield  {author} {\bibinfo {author} {\bibfnamefont {J.~C.}\ \bibnamefont
  {Slater}},\ }\href@noop {} {\bibfield  {journal} {\bibinfo  {journal} {Phys.
  Rev.}\ }\textbf {\bibinfo {volume} {81}},\ \bibinfo {pages} {385} (\bibinfo
  {year} {1951})}\BibitemShut {NoStop}%
\bibitem [{\citenamefont {{Le Bloas}}\ \emph {et~al.}(2011)\citenamefont {{Le
  Bloas}}, \citenamefont {Koh}, \citenamefont {Quentin}, \citenamefont
  {Bonneau},\ and\ \citenamefont {Ithnin}}]{LeBloas11_PRC84}%
  \BibitemOpen
  \bibfield  {author} {\bibinfo {author} {\bibfnamefont {J.}~\bibnamefont {{Le
  Bloas}}}, \bibinfo {author} {\bibfnamefont {M.~H.}\ \bibnamefont {Koh}},
  \bibinfo {author} {\bibfnamefont {P.}~\bibnamefont {Quentin}}, \bibinfo
  {author} {\bibfnamefont {L.}~\bibnamefont {Bonneau}}, \ and\ \bibinfo
  {author} {\bibfnamefont {J.~I.~A.}\ \bibnamefont {Ithnin}},\ }\href@noop {}
  {\bibfield  {journal} {\bibinfo  {journal} {Phys. Rev. C}\ }\textbf {\bibinfo
  {volume} {84}},\ \bibinfo {pages} {014310} (\bibinfo {year}
  {2011})}\BibitemShut {NoStop}%
\bibitem [{\citenamefont {Glöckle}(1983)}]{Glockle83}%
  \BibitemOpen
  \bibfield  {author} {\bibinfo {author} {\bibfnamefont {W.}~\bibnamefont
  {Glöckle}},\ }\href@noop {} {\emph {\bibinfo {title} {The Quantum Mechanical
  Few-Body Problem}}}\ (\bibinfo  {publisher} {Springer-Verlag},\ \bibinfo
  {year} {1983})\BibitemShut {NoStop}%
\bibitem [{\citenamefont {Hagen}\ \emph {et~al.}(2014)\citenamefont {Hagen},
  \citenamefont {Papenbrock}, \citenamefont {Hjorth-Jensen},\ and\
  \citenamefont {Dean}}]{Hagen14_RepProgPhys106}%
  \BibitemOpen
  \bibfield  {author} {\bibinfo {author} {\bibfnamefont {G.}~\bibnamefont
  {Hagen}}, \bibinfo {author} {\bibfnamefont {T.}~\bibnamefont {Papenbrock}},
  \bibinfo {author} {\bibfnamefont {M.}~\bibnamefont {Hjorth-Jensen}}, \ and\
  \bibinfo {author} {\bibfnamefont {D.~J.}\ \bibnamefont {Dean}},\ }\href@noop
  {} {\bibfield  {journal} {\bibinfo  {journal} {Rep. Prog. Phys.}\ }\textbf
  {\bibinfo {volume} {106}},\ \bibinfo {pages} {096302} (\bibinfo {year}
  {2014})}\BibitemShut {NoStop}%
\bibitem [{\citenamefont {Sun}\ \emph {et~al.}(2014)\citenamefont {Sun},
  \citenamefont {Bell}, \citenamefont {Hagen},\ and\ \citenamefont
  {Papenbrock}}]{Sun22_PRC106}%
  \BibitemOpen
  \bibfield  {author} {\bibinfo {author} {\bibfnamefont {Z.~H.}\ \bibnamefont
  {Sun}}, \bibinfo {author} {\bibfnamefont {C.~A.}\ \bibnamefont {Bell}},
  \bibinfo {author} {\bibfnamefont {G.}~\bibnamefont {Hagen}}, \ and\ \bibinfo
  {author} {\bibfnamefont {T.}~\bibnamefont {Papenbrock}},\ }\href@noop {}
  {\bibfield  {journal} {\bibinfo  {journal} {Phys. Rev. C}\ }\textbf {\bibinfo
  {volume} {106}},\ \bibinfo {pages} {L061302} (\bibinfo {year}
  {2014})}\BibitemShut {NoStop}%
\bibitem [{\citenamefont {Som{\`a}}\ \emph {et~al.}(2011)\citenamefont
  {Som{\`a}}, \citenamefont {Duguet},\ and\ \citenamefont
  {Barbieri}}]{Soma11_PRC84}%
  \BibitemOpen
  \bibfield  {author} {\bibinfo {author} {\bibfnamefont {V.}~\bibnamefont
  {Som{\`a}}}, \bibinfo {author} {\bibfnamefont {T.}~\bibnamefont {Duguet}}, \
  and\ \bibinfo {author} {\bibfnamefont {C.}~\bibnamefont {Barbieri}},\
  }\href@noop {} {\bibfield  {journal} {\bibinfo  {journal} {Phys. Rev. C}\
  }\textbf {\bibinfo {volume} {84}},\ \bibinfo {pages} {064317} (\bibinfo
  {year} {2011})}\BibitemShut {NoStop}%
\bibitem [{\citenamefont {Som{\`a}}\ \emph {et~al.}(2021)\citenamefont
  {Som{\`a}}, \citenamefont {Barbieri}, \citenamefont {Duguet},\ and\
  \citenamefont {Navr{\'a}til}}]{Soma21_EPJA57}%
  \BibitemOpen
  \bibfield  {author} {\bibinfo {author} {\bibfnamefont {V.}~\bibnamefont
  {Som{\`a}}}, \bibinfo {author} {\bibfnamefont {C.}~\bibnamefont {Barbieri}},
  \bibinfo {author} {\bibfnamefont {T.}~\bibnamefont {Duguet}}, \ and\ \bibinfo
  {author} {\bibfnamefont {P.}~\bibnamefont {Navr{\'a}til}},\ }\href@noop {}
  {\bibfield  {journal} {\bibinfo  {journal} {Eur. Phys. J. A}\ }\textbf
  {\bibinfo {volume} {57}},\ \bibinfo {pages} {135} (\bibinfo {year}
  {2021})}\BibitemShut {NoStop}%
\bibitem [{\citenamefont {Bally}\ and\ \citenamefont
  {Bender}(2021)}]{Bally21_PRC103}%
  \BibitemOpen
  \bibfield  {author} {\bibinfo {author} {\bibfnamefont {B.}~\bibnamefont
  {Bally}}\ and\ \bibinfo {author} {\bibfnamefont {M.}~\bibnamefont {Bender}},\
  }\href@noop {} {\bibfield  {journal} {\bibinfo  {journal} {Phys. Rev. C}\
  }\textbf {\bibinfo {volume} {103}},\ \bibinfo {pages} {024315} (\bibinfo
  {year} {2021})}\BibitemShut {NoStop}%
\bibitem [{\citenamefont {Tichai}\ \emph {et~al.}(2018)\citenamefont {Tichai},
  \citenamefont {Arthuis}, \citenamefont {Duguet}, \citenamefont {Hergert},\
  and\ \citenamefont {Som{\`a}}}]{Tichai18_PLB786}%
  \BibitemOpen
  \bibfield  {author} {\bibinfo {author} {\bibfnamefont {A.}~\bibnamefont
  {Tichai}}, \bibinfo {author} {\bibfnamefont {P.}~\bibnamefont {Arthuis}},
  \bibinfo {author} {\bibfnamefont {T.}~\bibnamefont {Duguet}}, \bibinfo
  {author} {\bibfnamefont {H.}~\bibnamefont {Hergert}}, \ and\ \bibinfo
  {author} {\bibfnamefont {V.}~\bibnamefont {Som{\`a}}},\ }\href@noop {}
  {\bibfield  {journal} {\bibinfo  {journal} {Phys. Lett. B}\ }\textbf
  {\bibinfo {volume} {786}},\ \bibinfo {pages} {195} (\bibinfo {year}
  {2018})}\BibitemShut {NoStop}%
\bibitem [{\citenamefont {Frosini}\ \emph
  {et~al.}(2022{\natexlab{a}})\citenamefont {Frosini}, \citenamefont {Duguet},
  \citenamefont {Ebran},\ and\ \citenamefont {Som{\`a}}}]{Frosini22_EPJA58_62}%
  \BibitemOpen
  \bibfield  {author} {\bibinfo {author} {\bibfnamefont {M.}~\bibnamefont
  {Frosini}}, \bibinfo {author} {\bibfnamefont {T.}~\bibnamefont {Duguet}},
  \bibinfo {author} {\bibfnamefont {J.-P.}\ \bibnamefont {Ebran}}, \ and\
  \bibinfo {author} {\bibfnamefont {V.}~\bibnamefont {Som{\`a}}},\ }\href@noop
  {} {\bibfield  {journal} {\bibinfo  {journal} {Eur. Phys. J. A}\ }\textbf
  {\bibinfo {volume} {58}},\ \bibinfo {pages} {62} (\bibinfo {year}
  {2022}{\natexlab{a}})}\BibitemShut {NoStop}%
\bibitem [{\citenamefont {Frosini}\ \emph
  {et~al.}(2022{\natexlab{b}})\citenamefont {Frosini}, \citenamefont {Duguet},
  \citenamefont {Ebran}, \citenamefont {B.Bally}, \citenamefont {T.Mongelli},
  \citenamefont {T.R.Rodriguez}, \citenamefont {R.Roth},\ and\ \citenamefont
  {Som{\`a}}}]{Frosini22_EPJA58_63}%
  \BibitemOpen
  \bibfield  {author} {\bibinfo {author} {\bibfnamefont {M.}~\bibnamefont
  {Frosini}}, \bibinfo {author} {\bibfnamefont {T.}~\bibnamefont {Duguet}},
  \bibinfo {author} {\bibfnamefont {J.-P.}\ \bibnamefont {Ebran}}, \bibinfo
  {author} {\bibnamefont {B.Bally}}, \bibinfo {author} {\bibnamefont
  {T.Mongelli}}, \bibinfo {author} {\bibnamefont {T.R.Rodriguez}}, \bibinfo
  {author} {\bibnamefont {R.Roth}}, \ and\ \bibinfo {author} {\bibfnamefont
  {V.}~\bibnamefont {Som{\`a}}},\ }\href@noop {} {\bibfield  {journal}
  {\bibinfo  {journal} {Eur. Phys. J. A}\ }\textbf {\bibinfo {volume} {58}},\
  \bibinfo {pages} {63} (\bibinfo {year} {2022}{\natexlab{b}})}\BibitemShut
  {NoStop}%
\bibitem [{\citenamefont {Frosini}\ \emph
  {et~al.}(2022{\natexlab{c}})\citenamefont {Frosini}, \citenamefont {Duguet},
  \citenamefont {Ebran}, \citenamefont {B.Bally}, \citenamefont {Hergert},
  \citenamefont {Rodriguez}, \citenamefont {Roth}, \citenamefont {Yao},\ and\
  \citenamefont {Som{\`a}}}]{Frosini22_EPJA58_64}%
  \BibitemOpen
  \bibfield  {author} {\bibinfo {author} {\bibfnamefont {M.}~\bibnamefont
  {Frosini}}, \bibinfo {author} {\bibfnamefont {T.}~\bibnamefont {Duguet}},
  \bibinfo {author} {\bibfnamefont {J.-P.}\ \bibnamefont {Ebran}}, \bibinfo
  {author} {\bibnamefont {B.Bally}}, \bibinfo {author} {\bibfnamefont
  {H.}~\bibnamefont {Hergert}}, \bibinfo {author} {\bibfnamefont
  {T.}~\bibnamefont {Rodriguez}}, \bibinfo {author} {\bibfnamefont
  {R.}~\bibnamefont {Roth}}, \bibinfo {author} {\bibfnamefont {J.~M.}\
  \bibnamefont {Yao}}, \ and\ \bibinfo {author} {\bibfnamefont
  {V.}~\bibnamefont {Som{\`a}}},\ }\href@noop {} {\bibfield  {journal}
  {\bibinfo  {journal} {Eur. Phys. J. A}\ }\textbf {\bibinfo {volume} {58}},\
  \bibinfo {pages} {64} (\bibinfo {year} {2022}{\natexlab{c}})}\BibitemShut
  {NoStop}%
\bibitem [{\citenamefont {Tsukiyama}\ \emph {et~al.}(2011)\citenamefont
  {Tsukiyama}, \citenamefont {Bogner},\ and\ \citenamefont
  {Schwenk}}]{Tsukiyama11_PRL106}%
  \BibitemOpen
  \bibfield  {author} {\bibinfo {author} {\bibfnamefont {K.}~\bibnamefont
  {Tsukiyama}}, \bibinfo {author} {\bibfnamefont {S.~K.}\ \bibnamefont
  {Bogner}}, \ and\ \bibinfo {author} {\bibfnamefont {A.}~\bibnamefont
  {Schwenk}},\ }\href@noop {} {\bibfield  {journal} {\bibinfo  {journal} {Phys.
  Rev. Lett.}\ }\textbf {\bibinfo {volume} {106}},\ \bibinfo {pages} {222502}
  (\bibinfo {year} {2011})}\BibitemShut {NoStop}%
\bibitem [{\citenamefont {Hergert}\ \emph {et~al.}(2016)\citenamefont
  {Hergert}, \citenamefont {Bogner}, \citenamefont {Morris}, \citenamefont
  {Schwenk},\ and\ \citenamefont {Tsukiyama}}]{Hergert16_PhysRep621}%
  \BibitemOpen
  \bibfield  {author} {\bibinfo {author} {\bibfnamefont {H.}~\bibnamefont
  {Hergert}}, \bibinfo {author} {\bibfnamefont {S.~K.}\ \bibnamefont {Bogner}},
  \bibinfo {author} {\bibfnamefont {T.~D.}\ \bibnamefont {Morris}}, \bibinfo
  {author} {\bibfnamefont {A.}~\bibnamefont {Schwenk}}, \ and\ \bibinfo
  {author} {\bibfnamefont {K.}~\bibnamefont {Tsukiyama}},\ }\href@noop {}
  {\bibfield  {journal} {\bibinfo  {journal} {Phys. Rep.}\ }\textbf {\bibinfo
  {volume} {621}},\ \bibinfo {pages} {165} (\bibinfo {year}
  {2016})}\BibitemShut {NoStop}%
\bibitem [{\citenamefont {Stroberg}\ \emph {et~al.}(2017)\citenamefont
  {Stroberg}, \citenamefont {Calci}, \citenamefont {Hergert}, \citenamefont
  {Holt}, \citenamefont {Bogner}, \citenamefont {Roth},\ and\ \citenamefont
  {Schwenk}}]{Stroberg17_PRL118}%
  \BibitemOpen
  \bibfield  {author} {\bibinfo {author} {\bibfnamefont {S.}~\bibnamefont
  {Stroberg}}, \bibinfo {author} {\bibfnamefont {A.}~\bibnamefont {Calci}},
  \bibinfo {author} {\bibfnamefont {H.}~\bibnamefont {Hergert}}, \bibinfo
  {author} {\bibfnamefont {J.}~\bibnamefont {Holt}}, \bibinfo {author}
  {\bibfnamefont {S.}~\bibnamefont {Bogner}}, \bibinfo {author} {\bibfnamefont
  {R.}~\bibnamefont {Roth}}, \ and\ \bibinfo {author} {\bibfnamefont
  {A.}~\bibnamefont {Schwenk}},\ }\href@noop {} {\bibfield  {journal} {\bibinfo
   {journal} {Phys. Rev. Lett.}\ }\textbf {\bibinfo {volume} {118}},\ \bibinfo
  {pages} {032502} (\bibinfo {year} {2017})}\BibitemShut {NoStop}%
\bibitem [{\citenamefont {Valderrama}\ \emph {et~al.}(2017)\citenamefont
  {Valderrama}, \citenamefont {{S\'anchez S\'anchez}}, \citenamefont {Yang},
  \citenamefont {Long}, \citenamefont {Carbonell},\ and\ \citenamefont {van
  Kolck}}]{PavonValderrama17_PRC95}%
  \BibitemOpen
  \bibfield  {author} {\bibinfo {author} {\bibfnamefont {M.~P.}\ \bibnamefont
  {Valderrama}}, \bibinfo {author} {\bibfnamefont {M.}~\bibnamefont {{S\'anchez
  S\'anchez}}}, \bibinfo {author} {\bibfnamefont {C.-J.}\ \bibnamefont {Yang}},
  \bibinfo {author} {\bibfnamefont {B.}~\bibnamefont {Long}}, \bibinfo {author}
  {\bibfnamefont {J.}~\bibnamefont {Carbonell}}, \ and\ \bibinfo {author}
  {\bibfnamefont {U.}~\bibnamefont {van Kolck}},\ }\href@noop {} {\bibfield
  {journal} {\bibinfo  {journal} {Phys. Rev. C}\ }\textbf {\bibinfo {volume}
  {95}},\ \bibinfo {pages} {054001} (\bibinfo {year} {2017})}\BibitemShut
  {NoStop}%
\bibitem [{\citenamefont {Birse}(2006)}]{Birse06_PRC74}%
  \BibitemOpen
  \bibfield  {author} {\bibinfo {author} {\bibfnamefont {M.}~\bibnamefont
  {Birse}},\ }\href@noop {} {\bibfield  {journal} {\bibinfo  {journal} {Phys.
  Rev. C}\ }\textbf {\bibinfo {volume} {74}},\ \bibinfo {pages} {014003}
  (\bibinfo {year} {2006})}\BibitemShut {NoStop}%
\bibitem [{\citenamefont {Wu}\ and\ \citenamefont {Long}(2019)}]{Wu19_PRC99}%
  \BibitemOpen
  \bibfield  {author} {\bibinfo {author} {\bibfnamefont {S.}~\bibnamefont
  {Wu}}\ and\ \bibinfo {author} {\bibfnamefont {B.}~\bibnamefont {Long}},\
  }\href@noop {} {\bibfield  {journal} {\bibinfo  {journal} {Phys. Rev. C}\
  }\textbf {\bibinfo {volume} {99}},\ \bibinfo {pages} {024003} (\bibinfo
  {year} {2019})}\BibitemShut {NoStop}%
\bibitem [{\citenamefont {{S\'anchez S\'anchez}}\ \emph
  {et~al.}(2020)\citenamefont {{S\'anchez S\'anchez}}, \citenamefont
  {Smirnova}, \citenamefont {Shirokov}, \citenamefont {Maris},\ and\
  \citenamefont {Vary}}]{Sanchez20_PRC102}%
  \BibitemOpen
  \bibfield  {author} {\bibinfo {author} {\bibfnamefont {M.}~\bibnamefont
  {{S\'anchez S\'anchez}}}, \bibinfo {author} {\bibfnamefont {N.~A.}\
  \bibnamefont {Smirnova}}, \bibinfo {author} {\bibfnamefont {A.~M.}\
  \bibnamefont {Shirokov}}, \bibinfo {author} {\bibfnamefont {P.}~\bibnamefont
  {Maris}}, \ and\ \bibinfo {author} {\bibfnamefont {J.~P.}\ \bibnamefont
  {Vary}},\ }\href@noop {} {\bibfield  {journal} {\bibinfo  {journal} {Phys.
  Rev. C}\ }\textbf {\bibinfo {volume} {102}},\ \bibinfo {pages} {024324}
  (\bibinfo {year} {2020})}\BibitemShut {NoStop}%
\bibitem [{\citenamefont {{S\'anchez S\'anchez}}\ \emph
  {et~al.}(2018)\citenamefont {{S\'anchez S\'anchez}}, \citenamefont {Yang},
  \citenamefont {Long},\ and\ \citenamefont {van Kolck}}]{Sanchez18_PRC97}%
  \BibitemOpen
  \bibfield  {author} {\bibinfo {author} {\bibfnamefont {M.}~\bibnamefont
  {{S\'anchez S\'anchez}}}, \bibinfo {author} {\bibfnamefont {C.-J.}\
  \bibnamefont {Yang}}, \bibinfo {author} {\bibfnamefont {B.}~\bibnamefont
  {Long}}, \ and\ \bibinfo {author} {\bibfnamefont {U.}~\bibnamefont {van
  Kolck}},\ }\href@noop {} {\bibfield  {journal} {\bibinfo  {journal} {Phys.
  Rev. C}\ }\textbf {\bibinfo {volume} {97}},\ \bibinfo {pages} {024001}
  (\bibinfo {year} {2018})}\BibitemShut {NoStop}%
\bibitem [{\citenamefont {Baye}(2015)}]{Baye15_PhysRep565}%
  \BibitemOpen
  \bibfield  {author} {\bibinfo {author} {\bibfnamefont {D.}~\bibnamefont
  {Baye}},\ }\href@noop {} {\bibfield  {journal} {\bibinfo  {journal} {Phys.
  Rep.}\ }\textbf {\bibinfo {volume} {565}},\ \bibinfo {pages} {1} (\bibinfo
  {year} {2015})}\BibitemShut {NoStop}%
\bibitem [{\citenamefont {H{\"u}ber}\ \emph {et~al.}(1997)\citenamefont
  {H{\"u}ber}, \citenamefont {Wita{\l}a}, \citenamefont {Nogga}, \citenamefont
  {Gl{\"o}ckle},\ and\ \citenamefont {Kamada}}]{Huber97_FBS22}%
  \BibitemOpen
  \bibfield  {author} {\bibinfo {author} {\bibfnamefont {D.}~\bibnamefont
  {H{\"u}ber}}, \bibinfo {author} {\bibfnamefont {H.}~\bibnamefont
  {Wita{\l}a}}, \bibinfo {author} {\bibfnamefont {A.}~\bibnamefont {Nogga}},
  \bibinfo {author} {\bibfnamefont {W.}~\bibnamefont {Gl{\"o}ckle}}, \ and\
  \bibinfo {author} {\bibfnamefont {H.}~\bibnamefont {Kamada}},\ }\href@noop {}
  {\bibfield  {journal} {\bibinfo  {journal} {Few-Body Systems}\ }\textbf
  {\bibinfo {volume} {22}},\ \bibinfo {pages} {107} (\bibinfo {year}
  {1997})}\BibitemShut {NoStop}%
\bibitem [{\citenamefont {Wendt}\ \emph {et~al.}(2011)\citenamefont {Wendt},
  \citenamefont {Furnstahl},\ and\ \citenamefont {Perry}}]{Wendt11_PRC83}%
  \BibitemOpen
  \bibfield  {author} {\bibinfo {author} {\bibfnamefont {K.~A.}\ \bibnamefont
  {Wendt}}, \bibinfo {author} {\bibfnamefont {R.~J.}\ \bibnamefont
  {Furnstahl}}, \ and\ \bibinfo {author} {\bibfnamefont {R.~J.}\ \bibnamefont
  {Perry}},\ }\href@noop {} {\bibfield  {journal} {\bibinfo  {journal} {Phys.
  Rev. C}\ }\textbf {\bibinfo {volume} {83}},\ \bibinfo {pages} {034005}
  (\bibinfo {year} {2011})}\BibitemShut {NoStop}%
\bibitem [{\citenamefont {van Dalen}\ and\ \citenamefont
  {M{\"u}ther}(2014)}]{vanDalen14_PRC90}%
  \BibitemOpen
  \bibfield  {author} {\bibinfo {author} {\bibfnamefont {E.~N.~E.}\
  \bibnamefont {van Dalen}}\ and\ \bibinfo {author} {\bibfnamefont
  {H.}~\bibnamefont {M{\"u}ther}},\ }\href@noop {} {\bibfield  {journal}
  {\bibinfo  {journal} {Phys. Rev. C}\ }\textbf {\bibinfo {volume} {90}},\
  \bibinfo {pages} {034312} (\bibinfo {year} {2014})}\BibitemShut {NoStop}%
\bibitem [{Dao(2019)}]{Dao20_PhD}%
  \BibitemOpen
  \href@noop {} {\bibfield  {journal} {\bibinfo  {journal} {Dao Duy Duc, PhD
  thesis, University of Bordeaux https://theses.hal.science/tel-02887649}\ }
  (\bibinfo {year} {2019})}\BibitemShut {NoStop}%
\bibitem [{\citenamefont {{Dao Duy Duc}}\ and\ \citenamefont
  {Bonneau}(2020)}]{Dao20_APhysPolBSupp13}%
  \BibitemOpen
  \bibfield  {author} {\bibinfo {author} {\bibnamefont {{Dao Duy Duc}}}\ and\
  \bibinfo {author} {\bibfnamefont {L.}~\bibnamefont {Bonneau}},\ }\href@noop
  {} {\bibfield  {journal} {\bibinfo  {journal} {Acta Phys. Pol. B Suppl.}\
  }\textbf {\bibinfo {volume} {13}},\ \bibinfo {pages} {405} (\bibinfo {year}
  {2020})}\BibitemShut {NoStop}%
\bibitem [{\citenamefont {Dudek}\ \emph {et~al.}(2022)\citenamefont {Dudek},
  \citenamefont {Dobaczewski}, \citenamefont {Dubray}, \citenamefont
  {G{\'o}{\'z}d{\'z}}, \citenamefont {Pangon},\ and\ \citenamefont
  {Schunck}}]{Dudek07_IJMPE16}%
  \BibitemOpen
  \bibfield  {author} {\bibinfo {author} {\bibfnamefont {J.}~\bibnamefont
  {Dudek}}, \bibinfo {author} {\bibfnamefont {J.}~\bibnamefont {Dobaczewski}},
  \bibinfo {author} {\bibfnamefont {N.}~\bibnamefont {Dubray}}, \bibinfo
  {author} {\bibfnamefont {A.}~\bibnamefont {G{\'o}{\'z}d{\'z}}}, \bibinfo
  {author} {\bibfnamefont {V.}~\bibnamefont {Pangon}}, \ and\ \bibinfo {author}
  {\bibfnamefont {N.}~\bibnamefont {Schunck}},\ }\href@noop {} {\bibfield
  {journal} {\bibinfo  {journal} {Eur. Phys. J. A}\ }\textbf {\bibinfo {volume}
  {58}},\ \bibinfo {pages} {64} (\bibinfo {year} {2022})}\BibitemShut {NoStop}%
\bibitem [{\citenamefont {Dudek}\ \emph {et~al.}(2010)\citenamefont {Dudek},
  \citenamefont {G{\'o}{\'z}d{\'z}}, \citenamefont {Mazurek},\ and\
  \citenamefont {Molique}}]{Dudek10_JPG37}%
  \BibitemOpen
  \bibfield  {author} {\bibinfo {author} {\bibfnamefont {J.}~\bibnamefont
  {Dudek}}, \bibinfo {author} {\bibfnamefont {A.}~\bibnamefont
  {G{\'o}{\'z}d{\'z}}}, \bibinfo {author} {\bibfnamefont {K.}~\bibnamefont
  {Mazurek}}, \ and\ \bibinfo {author} {\bibfnamefont {H.}~\bibnamefont
  {Molique}},\ }\href@noop {} {\bibfield  {journal} {\bibinfo  {journal} {J.
  Phys. G: Nucl. Part. Phys.}\ }\textbf {\bibinfo {volume} {37}},\ \bibinfo
  {pages} {064032} (\bibinfo {year} {2010})}\BibitemShut {NoStop}%
\bibitem [{\citenamefont {Ring}\ and\ \citenamefont
  {Schuck}(1980)}]{Ring-Schuck80}%
  \BibitemOpen
  \bibfield  {author} {\bibinfo {author} {\bibfnamefont {P.}~\bibnamefont
  {Ring}}\ and\ \bibinfo {author} {\bibfnamefont {P.}~\bibnamefont {Schuck}},\
  }\href@noop {} {\emph {\bibinfo {title} {The nuclear many-body problem}}}\
  (\bibinfo  {publisher} {Springer-Verlag},\ \bibinfo {year}
  {1980})\BibitemShut {NoStop}%
\bibitem [{\citenamefont {Varshalovich}\ \emph {et~al.}(1988)\citenamefont
  {Varshalovich}, \citenamefont {Moskalev},\ and\ \citenamefont
  {Kherkonskii}}]{Varshalovich88}%
  \BibitemOpen
  \bibfield  {author} {\bibinfo {author} {\bibfnamefont {D.~A.}\ \bibnamefont
  {Varshalovich}}, \bibinfo {author} {\bibfnamefont {A.~N.}\ \bibnamefont
  {Moskalev}}, \ and\ \bibinfo {author} {\bibfnamefont {V.~K.}\ \bibnamefont
  {Kherkonskii}},\ }\href@noop {} {\emph {\bibinfo {title} {Quantum Theory of
  Angular Momentum}}}\ (\bibinfo  {publisher} {World Scientific, Singapore},\
  \bibinfo {year} {1988})\BibitemShut {NoStop}%
\bibitem [{\citenamefont {Stoks}\ \emph {et~al.}(1993)\citenamefont {Stoks},
  \citenamefont {Klomp}, \citenamefont {Rentmeester},\ and\ \citenamefont
  {de~Swart}}]{PWA93}%
  \BibitemOpen
  \bibfield  {author} {\bibinfo {author} {\bibfnamefont {V.~G.~J.}\
  \bibnamefont {Stoks}}, \bibinfo {author} {\bibfnamefont {R.~A.~M.}\
  \bibnamefont {Klomp}}, \bibinfo {author} {\bibfnamefont {M.~C.~M.}\
  \bibnamefont {Rentmeester}}, \ and\ \bibinfo {author} {\bibfnamefont {J.~J.}\
  \bibnamefont {de~Swart}},\ }\href@noop {} {\bibfield  {journal} {\bibinfo
  {journal} {Phys. Rev. C}\ }\textbf {\bibinfo {volume} {48}},\ \bibinfo
  {pages} {792} (\bibinfo {year} {1993})}\BibitemShut {NoStop}%
\bibitem [{\citenamefont {Wong}\ and\ \citenamefont
  {Clement}(1972)}]{Wong72_NPA183}%
  \BibitemOpen
  \bibfield  {author} {\bibinfo {author} {\bibfnamefont {C.~W.}\ \bibnamefont
  {Wong}}\ and\ \bibinfo {author} {\bibfnamefont {D.~M.}\ \bibnamefont
  {Clement}},\ }\href@noop {} {\bibfield  {journal} {\bibinfo  {journal} {Nucl.
  Phys. A}\ }\textbf {\bibinfo {volume} {183}},\ \bibinfo {pages} {210}
  (\bibinfo {year} {1972})}\BibitemShut {NoStop}%
\bibitem [{\citenamefont {Moshinsky}(1959)}]{Moshinsky59_NP13}%
  \BibitemOpen
  \bibfield  {author} {\bibinfo {author} {\bibfnamefont {M.}~\bibnamefont
  {Moshinsky}},\ }\href@noop {} {\bibfield  {journal} {\bibinfo  {journal}
  {Nucl. Phys.}\ }\textbf {\bibinfo {volume} {13}},\ \bibinfo {pages} {104}
  (\bibinfo {year} {1959})}\BibitemShut {NoStop}%
\end{thebibliography}
